# Main Belt Binary Asteroidal Systems With Eccentric Mutual Orbits[*]

F. Marchis[a,b,c], P. Descamps[b], J. Berthier[b], D. Hestroffer[b], F. Vachier[b], M. Baek[c], A. Harris[d], D. Nesvorny[e]

[a] University of California at Berkeley, Department of Astronomy, 601 Campbell Hall, Berkeley, CA 94720, USA
[b] Institut de Mécanique Céleste et de Calcul des Éphémérides, Observatoire de Paris, 75014 Paris, France
c. SETI Institute, Carl Sagan Center, 515 N. Whismann Road, Mountain View CA 94043, USA
d. DLR Institute of Planetary Research, Rutherfordstrasse 2, 12489 Berlin, Germany
e. Southwest Research Institute, 1050 Walnut Street, Suite 400, Boulder, CO 80302, USA

Pages: 60
Tables: 7
Figures: 4

**Proposed running head:** eccentric mutual orbits of binary asteroidal systems

**Editorial correspondence to:**
Franck Marchis
601 Campbell Hall
Berkeley CA 94720
USA
Phone: +1 510 642 3958
Fax: +1 510 642 3411
Email: fmarchis@berkeley.edu

## ABSTRACT

Using 8m-10m class telescopes and their Adaptive Optics (AO) systems, we conducted a long-term adaptive optics campaign initiated in 2003 focusing on four binary asteroid systems: (130) Elektra, (283) Emma, (379) Huenna, and (3749) Balam. The analysis of these data confirms the presence of their asteroidal satellite. We did not detect any additional satellite around these systems even though we have the capability of detecting a loosely-bound fragment (located at $1/4 \times R_{Hill}$) ~40 times smaller in diameter than the primary. The orbits derived for their satellites display significant eccentricity, ranging from 0.1 to 0.9, suggesting a different origin. Based on AO size estimate, we show that (130) Elektra and (283) Emma, G-type and P-type asteroids respectively, have a significant porosity (30-60% considering CI-CO meteorites as analogs) and their satellite's eccentricities ($e$~0.1) are possibly due to excitation by tidal effects. (379) Huenna and (3749) Balam, two loosely bound binary systems, are most likely formed by mutual capture. (3749) Balam's possible high bulk density is similar to (433) Eros, another S-type asteroid, and should be poorly fractured as well. (379) Huenna seems to display both characteristics: the moonlet orbits far away from the primary in term of stability (20% × $R_{Hill}$), but the primary's porosity is significant (30-60%).

# 1. Introduction

It was only when the first images of the asteroid (243) Ida captured by the Galileo spacecraft revealed the presence of a small satellite named Dactyl, that the existence of binary asteroid suggested by Andre (1901) and discussed in Van Flandern *et al.* (1979) was unambiguously confirmed. The advent of high angular resolution imaging provided by instruments such as ground-based telescopes equipped with adaptive optics (AO) systems, and also by the Hubble Space Telescope, permitted the discovery of new visual binary asteroids (Noll, 2006; Richardson and Walsh, 2006). Radar observations of Near Earth Asteroids during a close passage with Earth also revealed that binary systems are common in this population (Margot et al., 2002). At the time of writing, more than sixty systems have been imaged, but the number of suspected binary asteroids is significantly higher (~145) since many of them display mutual event signatures (Behrend *et al.* 2006, Descamps *et al.* 2007) and/or multi-period components (Pravec and Harris, 2007) in their lightcurves. Despite recent simulations involving catastrophic collisions (Durda *et al.* 2004), fission via the YORP effect (Cuk *et al.* 2005), and split due to tidal effect with a major planet (Walsh and Richardson, 2006) among others, the formation of most of these multiple asteroid systems is not yet understood. Insights into these binary systems, such as the orbital parameters of the satellite, the size and shape of the components of the system, the nature of their surface, their bulk density and distribution of materials in their interior could provide a better understanding of how these multiple asteroidal systems formed.

Over the past few years, our group has focused its attention on binaries located in the main-belt which have been discovered visually. We initiated an intensive campaign of observations from 2003 through 2006 combining the adaptive optics high-resolution capabilities of various 8m-class telescopes (UT4 of the Very Large Telescope, W.M. Keck-II and Gemini-North) equipped with Adaptive Optics (AO) systems that allow us to resolve the binary system. This project is part of the LAOSA (Large Adaptive Optics Survey of Asteroids, Marchis *et al.* 2006c), which aims to discover binary asteroids and study their characteristics using high angular capabilities provided by large aperture telescopes with AO systems. We have separately published (Descamps *et al.* 2007) a complete analysis of the orbit and size and shape of the components of (90) Antiope, which is a doublet binary system (i. e. composed of two similarly-sized components). In this work, we focus on binary asteroidal systems with smaller satellites (also called "moonlet companions"). In Section 2 of this article, we present the resolved AO observations of four binary systems, (130) Elektra, (283) Emma, (379) Huenna, and (3749) Balam. Section 3 describes how we derive the orbits of these systems which display significant eccentricities. In Section 4 we estimate the average diameter, shape, and bulk density of the (130) Elektra and (283) Emma systems using direct resolved observations of the primary. An estimate of the bulk density and the porosity of (379) Huenna and (3749) Balam are described in the next section. Finally, we discuss the origin of these systems based on their measured characteristics in Section 5.

## 2. Adaptive optics observations

**2.1 Collected data and basic data reduction**

The concept of adaptive optics was proposed by Babcock (1953), but it was not until the end of the 1980's that the first prototypes were developed independently by several groups based in the United States and France. The AO systems provide in real-time an image with an angular resolution close to the diffraction limit of the telescope. Because of technological limitations, linked to the way the wavefront is analyzed, most of the AO systems procure a correction that is only partial and slightly variable in time in the NIR (1-5μm). Several AO systems are now available on 8m-class telescopes, such as Keck-10m II, Gemini-8m North both at Mauna Kea (Hawaii, USA) and the UT4-Yepun of the Very Large Telescope observatory at Paranal (Chile). These systems provide a stable correction in K-band (2.2 μm), with an angular resolution close to the diffraction limit of the telescope; 60 milli-arcsec (*mas*) for the Gemini and the VLT, and 50 mas for the Keck under good exterior seeing conditions (<0.8") on targets brighter than the 13[th] magnitude in the visible range.

Since 1998, several binary asteroid systems were discovered using various AO systems. The first one was Petit-Prince, a companion of 45 Eugenia, imaged with PUEO an AO mounted on the Canada-France-Hawaii 3.6m-telescope (Merline *et al.* 1999). Since then ~14 main-belt binary asteroids have been discovered using this technique on 8m-10m class telescopes.

In 2004, we initiated a large campaign of observations using the UT4 of the Very Large Telescope (VLT) of the European Southern Observatory and its AO system called NAOS (Nasmyth Adaptive Optics System). The observations were recorded in direct imaging using the CONICA near-infrared camera equipped with an ALADDIN2 1024×1024 pixel InSb array of 27 μm pixels. Most of the data were recorded with the S13 camera (13.27 *mas*/pixel scale) in Ks band (central wavelength 2.18 μm and bandwidth of 0.35 μm). NACO, which stands for NAOS-CONICA, provides the best angular correction in this wavelength range (Lenzen *et al.* 2003, Rousset *et al.* 2003). Approximately 70 hours of observations were allocated to this program in service observing. In 2005 and 2006, we continued this program using the Gemini North telescope and its recently commissioned AO system called ALTAIR (Herriot *et al.* 2000). ALTAIR feeds NIRI (Hodapp *et al.* 2003), a near-infrared instrument. NIRI equipped with a 1024 x 1024 pixel ALADIN InSB array sensitive from 1 to 5 microns was used in imaging mode along with the f/32 cameras providing a pixel scale of 22 mas. Twelve hours of observations were recorded in queue scheduling under median seeing conditions of ~1.0" with this instrument. On a few occasions during this campaign, complementary Ks band observations taken with the Keck-II AO and its Near-InfraRed Camera (NIRC2) were added to our analysis. We also included in the LAOSA database (Marchis et al., 2006b) observations of small solar system bodies that we could retrieve from Gemini-North and VLT archive centers corresponding to ~1100 observations of ~360 main-belt and ~50 Trojan asteroids.

    The basic data processing (sky substraction, bad-pixel removal, and flat-field correction) applied on all these raw data was performed using the *eclipse* data reduction package (Devillard, 1997). Successive frames taken over a time span of less than 6 min, were combined into one single average image after applying an accurate shift-and-add process through the *Jitter* pipeline offered in the same package. Data processing with this software on such high S/N data (>1000) is relatively straightforward, since the centroid position on each frame can be accurately measured by a Gaussian fit. The final image is obtained by stacking the set of cross-correlated individual frames.

**2.2 Targets**

This work describes the analysis of 4 main-belt minor planets already known to have a satellite: (130) Elektra, (283) Emma, (379) Huenna, and (3749) Balam. S/2003 (130) 1, a provisional name for the companion of the G-type asteroid (130) Elektra (Tholen *et al.* 1989) with a diameter estimated to 182 km (Tedesco *et al.* 2002) was first seen by Merline *et al.* (2003b) using the Keck II AO system in August 2003. One month earlier using the same instrument, the same group had reported the discovery of a moonlet companion temporarily named S/2003 (283) 1 of (283) Emma (Merline *et al.* 2003a). The taxonomic classification of this 148 km diameter asteroid is unclear. Tholen and Barucci (1989) placed it in the X-type family. In the S3OS2 survey (Lazzaro et al., 2004), this asteroid is classified as a C-type. In August 2003, the binary nature of (379) Huenna was revealed using the Keck-II AO system by Margot (2003). The IRAS radiometric diameter of this C-type asteroid (Bus and Binzel, 2002) is estimated to be 92 km (Tedesco *et al.* 2002). (3749) Balam's companion was discovered in February 2002 using the Hokupa'a AO mounted on the Gemini-North telescope by Merline *et al.* (2002a). Table 1 summarizes the known characteristics of these minor planets extracted from various published sources.

For all these systems, the orbital parameters of the companion orbit were previously unknown or poorly defined. The main motivation of this work was to obtain an accurate knowledge of their orbit that allow us to calculate directly the mass of the system from the Kepler's third law, the characteristics of the moonlet and the primary, and eventually the bulk density and porosity of the primary. Table 2a and Table 2b contain the observing log of all reduced observations of these binary systems extracted from the LAOSA database. The number of observations is variable between asteroids and between AO instruments. For instance, because of their faintness ($m_v$~16), (3749) Balam and (379) Huenna were observed only 16 and 33 times respectively with the VLT telescope, the only one equipped with an AO system able to provide a partial correction on such faint targets. (130) Elektra and (283) Emma have a predicted brightness magnitude in the visible ranging from 11.2 to 15.1, making these targets accessible to the Gemini and Keck AOs which are limited to 14-15$^{th}$ magnitude.

**Material:**

**Table 1: Characteristics of the studied minor planets**

**Table 2a,b: Observing conditions of AO observations**

**2.3 Search for moonlet companions**

Searching for a point source around a bright asteroid is not a trivial task even with an AO system. The Point Spread Function (PSF) of an AO system is composed of a coherent peak surrounded by a halo in which speckle patterns are also present. Because these speckle artifacts are variable in time and have an angular size corresponding to the diffraction-limit of the telescope, as well as a faint intensity ($\Delta m>7$), they could be easily mistaken for moonlet satellites. Additionally the presence of a continuous halo around the primary limits the signal-to-noise ratio on the detected moonlet and thus the accuracy on its position and its photometry.

We have developed and described in a previous work (Marchis *et al.* 2006b) a method to reduce the halo effect and estimate the upper limit of detection for AO observations. We applied this algorithm to all observations of the four binary systems. Table 3a and Table 3b summarize the characteristics of their synthesized 2-σ detection profiles. As previously shown in Marchis *et al.* (2006b), the 3 parameters ($\alpha$, $\Delta m_{lim}$, $r_{lim}$) that characterize the synthesized detection profile are quite variable. For r > $r_{lim}$ the detection profile (~$\Delta m_{lim}$) is roughly constant on the image. These parameters depend on parameters such as the seeing conditions, the airmass, the brightness of the object, the total integration time during the observations, but also the telescope and the design of its AO system. For instance, in the case of (130) Elektra, $\alpha$ varies from -9.1 to -3.2 and $\Delta m_{lim}$ from -9.4 to -5.9. This disparity in the detection profile can be directly translated into the minimum diameter size (5 to 34 km or 3 to 12 km) for a moon to be detected if located at 2/100 × $R_{Hill}$ or 1/4 × $R_{Hill}$, respectively.

In Figure 1a, 1b, 1c, and 1d we detail each step of the detection curve profile analysis for one observation of each asteroid. Subtracting the azimuthally averaged function improved the detection of the moonlet. The characteristics of the synthesized detection profile are also displayed. The two regimes separated by $r_{lim}$ are obvious on these detection profiles. We detect the companion unambiguously in 10 out of 44 observations of (130) Elektra, 25 out of 38 for (283) Emma, 25 out of 33 for (379) Huenna, and 7 out of 16 for (3749) Balam. The low detection rate for (130) Elektra is mostly due to poor seeing conditions during the Gemini run in April 2006 together with an edge-on appearance of the orbit. The moonlet was therefore located too close to the primary and its flux was lost in the halo due to the uncorrected residual phase of the AO. None of our observations show the presence of another moonlet around these binary systems, even though we had the capability of detecting a loosely-bound fragment (located at 1/4 × $R_{Hill}$) ~40 times smaller in diameter than the primary for (130) Elektra and (283) Emma. (87) Sylvia with its two moons Romulus and Remus (Marchis *et al.* 2005a) was the only known multiple system located in the main-belt, until Marchis et al. (2007b) announced in March 2007 the discovery of a second smaller and closer moonlet around (45) Eugenia.

**Material:**

**Table 3a and 3b**

**Fig 1a,1b,1c,1d**

**2.4 Size and shape of (130) Elektra's primary**

With an average angular size of 120 mas measured directly on our AO images, the (130) Elektra primary is resolved on 16 collected observations (Table 4a). The recording of punctual sources, such as unresolved stars, indicate that a typical AO PSF is characterized by a peak of coherent light (that defines the angular resolution) surrounded by a halo produced by the uncorrected residual phase. It is possible to improve the sharpness on our collected images by applying an *a posteriori* deconvolution numerical process. We developed AIDA, which is described thoroughly in Hom *et al.* (2007) and tested extensively in Marchis *et al.* (2006b) on asteroid-type images. To improve the sharpness of the images, AIDA algorithm (see Fig. 2) was applied. We used as an approximation of the Point Spread Function (PSF), an observation of a star or an unresolved asteroid recorded on the same night or run. The size and shape of the primary were approximated fitting them by an ellipse, of which major-axes and orientation are listed in Table 4a. With this technique and using the Keck AO (Dec. 7 2003) data that have the best angular resolution, our diameter estimate is accurate to 3%, corresponding to ~4 km for Elektra images. The errors are significantly higher (7-15 km) for observations taken in 2005-2006 when the asteroid was located at more than 2.5 AU from Earth.

We compared the apparent shape of Elektra's primary with the model developed by lightcurve inversion (Durech *et al.* 2007). As mentioned by Marchis et al. (2006b), the pole solution, (pole I with $\lambda = 68°$, $\beta=-88°$ in ECJ2000) and a spin period 5.2247 h seem to reproduce the geometry of the resolved image of Elektra taken in Dec 2003. The resolved images provided by AO permit to remove the ambiguity between two pole solutions which appears for asteroid orbiting close to the ecliptic. To check the validity of this pole solution, we display in Fig. 2 the projected shape of Elektra, the appearance from the model pole I and the almost symmetrical solution (pole II: $\lambda = 277°$, $\beta=85°$ in ECJ2000 which corresponds to the pole solution of the moonlet orbit (Section 3). The apparent orientation of Elektra generated with the pole I solution is remarquably similar to the observations (see Table 4a). The observations recorded on Jan. 5, 2004 and Jan. 15, 2005 are clearly different in appearance than the pole II model. A quantitative analysis indicates that pole II image orientations are shifted by 30º in average whereas pole I image orientations are closer to the observation with a 10º shift in average. This comparison implies that the pole I solution chosen in Durech *et al.* (2007) is a good approximation. It also signifies that the moonlet is orbiting around the primary in the opposite direction to the primary spin. This important result needs to confirm by carefully analyzing the combination of our AO data with the lightcurve photometric measurements. Since the sense of revolution of the moonlet around the primary with respect to the primary spin does not have consequences on the mass, and density determination discussed in the rest of this article, we will only state here this interesting possibility. The average diameter estimated on our AO observations is 215 ± 15 km, which is 16% larger than IRAS radiometric diameter by Tedesco *et al.* (2002). The tendency of IRAS radiometric measurements to underestimate the diameter of large and elongated asteroids has already been noted for various main-belt asteroids, such as (87) Sylvia (Marchis *et al.* 2005a) and (130) Elektra (Marchis *et al.* 2006b).

**Material:**

**Table 4a,4b**

**Figure 2 & 3**

**2.5 Size and shape of (283) Emma's primary**

The shape and size of (283) Emma's primary were measured using the same technique detailed in Section 2.4 for (130) Elektra. Table 4b detailed the orientation and size ratio after fitting by an ellipsoid. The angular size of this asteroid is slightly less than twice the angular resolution of an 8m-telescope leading to uncertainty of 7%. The projected shape is very close to an ellipse suggesting that the asteroid has a shape close to a perfect ellipsoid. The average diameter extracted from our observations is 160 ± 10 km, with an average a/b = 1.2. This measurement is in agreement with the only published lightcurve by Stanzel *et al.* (1978) reporting a spin period of 6.888 h and a regular lightcurve with a magnitude range ~0.3. Radiometric diameter reported by Tedesco *et al.* (2002) based on two sightings is $D_{STM}=148 \pm 5$ km.

Additional lightcurve observations are encouraged for this target to help to construct its 3D-shape model. It could be refined taking into consideration these resolved AO observations. This is an interesting main-belt asteroid since it was known to be member of the Eos collisonal family (Zappala et al., 1995), but recent work published by Nesvorny *et al.* (2006) suggested that in fact it is the largest member of its own collisional family.

**2.6 Astrometric positions and photometric measurements on the satellite**

The positions of the satellite with respect to its primary are measured as described in Marchis *et al.* (2005b). On each individual reduced image we estimate the position of the center of light of the primary and the secondary using a two-dimensional Moffat-Gauss fit profile (Descamps et al., 2002). The background around the satellite, introduced by residual errors in the AO correction, is modeled by an inclined quadratic surface. The plate scale used for each instrument was the one measured during their commissioning. In the case of Keck/NIRC2, although its platescale is poorly known (5% accuracy corresponding to 0.5 mas per pixel, so up to 4 mas in the case of 130 Elektra moonlet which is at 0.7"), it is of the same order than the accuracy of our fitted positions (~5 mas).

The astrometric positions relative to the primary in arcsec are labeled X and Y in Table 5a-5d. They correspond to the projected separation on the celestial sphere between the primary and the satellite: $X = \delta RA \times \cos(<DEC>)$ and $Y = \delta DEC$ with X positive when the satellite is located on the astronomical East of the primary and Y positive when it is locate d North.

From the Moffat-Gauss profile we also estimate the relative integrated flux between the moonlet and the primary. In the case of (130) Elektra and (283) Emma, the primary is also directly resolved on the AO images (see Sections 2.4 and 2.5 respectively). Taking the integrated flux of the Moffat-Gauss profile on the primary ($\Phi_{primary} = \int F_{primary}$) and secondary ($\Phi_{sat} = \int F_{sat}$), using the average diameter measured on the primary ($D_{av}$), and assuming the same albedo for the satellite and the primary, we derived the diameter of the secondary ($D_{sat}$) using the relation

$$D_{sat} = D_{av} \times (\Phi_{sat}/\Phi_{primary})^{1/2} \quad (1)$$

The satellite diameters of the (379) Huenna and (3749) Balam, whose primaries are not resolved, can be derived by comparing the peak-to-peak ratio through the relation

$$D_{sat} = D_{av} \times (\max(F_{sat})/\max(F_{primary}))^{1/2} \quad (2)$$

The diameter of each moonlet is given in Table 1. The Elektra, Emma, and Huenna systems are characterized by a small satellite companion (1/16- 1/30 the diameter of the primary) whereas Balam's satellite is half the diameter of its primary. The uncertainties in the size measurements of the moonlets are large (up to 60% in the case of Emma) because of the difficulty in extracting the weak flux of the moonlet orbiting close to the primary asteroid. The residual intensity due to the noise in the AO loop produced a halo around the primary, the intensity of which varies both temporally and spatially. However, it is also possible that this flux variation observed on the moonlet is partially real due to an irregular shape of the satellite. The availability of better AO systems (Next Generation of AO at Keck, GPI at Gemini) with better and more stable Strehl Ratio should reduce the halo intensity and variation, and allow us to improve the size estimate of the moonlet in the future.

> **Material:**
>
> **Table 5a,b,c,d:**
>
> **Figure 3**

## 3. Orbit determination

### 3.1 Method

Using these accurate astrometric data, we can estimate the true orbit of these systems. Descamps (2005) developed the Binary Orbit Fit (BOF) algorithm for this task based on the geometrical fitting of an apparent orbit and its dynamical evolution due to precession. As a first step, the relative positions of the satellite, i.e. the projected apparent positions on the plane of the sky, over a short period of time (~1 month), are used to estimate two apparent mirror orbits. These positions must be chosen in a way that they are spread out along the orbit. Next, we used the least square fitting of all observed positions to refine the complete set of orbital parameters and determine the best-fitting and unique solution for the pole of the orbit by varying $J_2$ (corresponding to the precession of the apsidal and nodal lines due to the oblatness of the primary), introducing an inclination for the satellite orbit if necessary, and correcting for light time and changes of viewing geometry due to parallax effects. Figures 3a-3d display the apparent orbit of the four studied binary systems. Their orbital parameters are summarized in Table 6. Our results were validated independently with the StatOrbit algorithm developed by Hestroffer *et al.* (2005) which uses both a geometrical and statistical approach. We are therefore confident that our orbital elements are well defined. BOF has already been used to estimate the orbits of various other binary asteroids (Marchis *et al.* 2005ab, Marchis *et al.* 2006a). In November 2006, a group of astronomers reported the observations of a secondary stellar occultation event by Linus, companion of (22) Kalliope. The event was detected very close to the position predicted by our model (Soma *et al.* 2006), providing independent validation of our orbit solution.

**3.2 Orbital parameters comparison**

The orbital parameters of Elektra, Emma, and Huenna could be estimated thanks to the good distribution of the positions along the orbit (see Fig. 3). In the case of (3749) Balam, our analysis had to take into account the fact that the moonlet was not detected in various observations taken on Nov. 15 and Nov 16, and was barely detectable (because it was near the primary) on Nov. 14 and Nov. 22 2004 (the same was true for two nights of observations on July 15 and July 16 2003). These additional but imprecise positions were necessary to extract the orbital parameters of Balam's satellite.

Our fitted elements for the orbits of the satellites are shown in Table 6. The apparent projected orbit and a display of the residuals on the positions for each binary system are displayed in Fig. 3a-3d. Using Kepler's third law (Kepler, 1609), it is possible to compute the mass of the system (Table 7). The 1-$\sigma$ uncertainties on the mass (~7-10%) are dominated by the precision of the semi-major axis measurement (2-4%).

> **Material**
>
> **Table 6 & 7**

The four binary systems display similarities and obvious differences. In comparison with previously published orbits of main-belt binary asteroids with moonlet companions ((22) Kalliope in Marchis *et al.* 2003; (45) Eugenia in Merline *et al.* 1999; (87) Sylvia in Marchis *et al.* 2005a; (121) Hermione in Marchis *et al.* (2005b)), these satellites have significantly eccentric orbits around their primaries.

Although its orbit is not well defined, S/2001(3749)1 is clearly the most eccentric. The best-fitted solution corresponds to an orbit with $e\sim0.9$, which is possibly the highest eccentricity of any moon in the Solar System and is higher, for instance, than that of the TNO 1998WW$_{31}$ (e~0.8, see Veillet *et al.* 2002). The orbits of (130) Elektra and (283) Emma companions are slightly eccentric ($e\sim0.1$) whereas the moon of (349) Huenna has an intermediate eccentricity ($e\sim0.3$).

Because the masses and the relative sizes of the components of the system (assuming the same albedo) are well constrained, it is possible to calculate accurately the Hill sphere radius around the primary (Table 7). The moonlets of (130) Elektra and (283) Emma orbit well-inside the Hill sphere of the primary (2% and 5% respectively) like most of the known binary systems, including (22) Kalliope, (87) Sylvia, and (121) Hermione (Marchis *et al.* 2003, Marchis *et al.* 2005ab). With a semi-major axis of half the Hill radius, the (349) Huenna and (3749) Balam satellites are both loosely-bound binary asteroids. Based on an incomplete orbit (*e* unknown, *a* and *P* approximated), Merline *et al.* (2002b) suggested this possibility for (3749) Balam companion. Our orbital measurements confirm unambiguously the existence of such a system in the main-belt. These differences in eccentricity and in semi-major axis suggest a different formation scenario for these binary systems.

4. **Internal structure: bulk density and porosity**

Using the mass from the analysis of the orbit ($M_{system}$) as well as the average radius estimated from radiometric IRAS measurements, we should be able to derive the bulk density of these binary systems. Tedesco *et al.* (2002) published an analysis of the IRAS data containing the average diameter of ~2200 minor planets. They used a simple thermal model based on spherical geometry called the Standard Thermal Model (STM) designed for large asteroids with low thermal inertia and/or slow rotation. Harris (1998) considered a modified approach with a model, called NEATM, developed specifically for Near-Earth asteroids including fast rotator with significant thermal inertia, but also valid for asteroids in general. With NEATM, the model temperature distribution is adjusted via the beaming parameter $\eta$ to force consistency with the observed apparent color temperature of the asteroid, which depends on thermal inertia, surface roughness, and spin vector. In the STM, the value of $\eta$ is kept constant (0.756) to take into account the surface roughness at low phase angle (see Harris, 2006 and references therein). The STM can give erroneous results for asteroids with thermal inertia and/or surface roughness different from those of the asteroids Ceres and Pallas against which it was calibrated (Lebofsky *et al.* 1986). Table 1 contains the average radius estimated using both methods based on the IRAS measurements for three asteroids with reported IRAS observations. The average radii vary significantly between both methods leading to a possible variation in their bulk density up to 20%.

In the case of (130) Elektra and (283) Emma, the angular resolution provided by the AO observations has been useful to estimate directly an approximation of the primary diameter. Table 7 summarizes the bulk density measurements using these diameter estimates.

**4.1 Density of (130) Elektra, a G-type asteroid.**

Tholen and Barucci (1989) classified (130) Elektra as a G-type asteroid, a sub-class of the C class, with low albedo and a strong absorption band at 0.4 μm. Based on $D_{STM}$ = 182 km or $D_{NEATM}$=196 km, we derived a bulk density of 2.1 or 1.7 g/cm$^3$ (± 0.3) respectively (based on 7 IRAS sightings). This density measurement is very close to the bulk density of (1) Ceres, another G-type asteroid, which was inferred by Thomas *et al.* (2005) from the ellipsoidal shape of this large asteroid. CI-CM carbonaceous meteorites (Britt and Consolmagno, 2003) are the best candidates for meteorite analogs in terms of bulk density (~2.1 g/cm3), suggesting no macro-porosity in the interior of the primary. For completeness, (130) Elektra is classified as a Ch-type in the SMASSII taxonomy (Bus and Binzel, 2002). The spectrum shows a relatively strong 0.7-micron phyllo-silicate absorption band. In this case, spectrally different than (1) Ceres, (130) Elektra is most analogous to CM-chondrites ($\rho_{average}$ = 2.12 g/cm$^3$, Britt and Consolmagno, 2003). This IRAS bulk density measurement suggests an absence of macro-porosity in the interior of the primary.

Considering the diameter estimate from our AO data ($D_{AO}$=215±15 km), we obtained a significantly lower bulk density (~1.3 ± 0.3 g/cm$^3$), which is of the same order as the measured bulk densities of the multiple C-type asteroids, including (45) Eugenia (Merline *et al.* 1999), (87) Sylvia (Marchis *et al.* 2005a), (90) Antiope (Descamps *et al.* 2007) and (121) Hermione (Marchis *et al.* 2005b). Considering carbonaceous meteorites as analogs for this asteroid, we derive a significant macro-porosity (30-50%) suggesting a possible rubble-pile interior. Independent measurements of Elektra primary diameter ($D_{Spitzer}$=202 ± 20 km) based on Spitzer IRS spectral data (J. Emery, personal communication) support the larger NEATM and AO diameter estimates.

**4.2 Low bulk density of (283) Emma, a P-type asteroid?**

(283) Emma is classified as a X-type by Tholen and Barucci (1989) a large class containing the E, M, and P spectral classes. The degeneracy between these taxonomic classes can be removed given the low albedo ($p_v$ = 0.03) reported by Tedesco *et al.* (2002), suggesting that this is a P-type asteroid. From the analysis of the orbit and the IRAS diameter estimate (based on 2 sightings) we derive a bulk density $\rho$ = 0.7-1.0 g/cm$^3$, similar to that of (617) Patroclus, a P-type Trojan (Marchis *et al.* 2006a). Considering CI carbonaceous chondrites as meteorite analogs with a bulk density of 2.11 g/cm$^3$ and a micro-porosity of 10% in Britt and Consolmagno (2003), we derived a significant macro-porosity (~50-60%) suggesting a rubble-pile internal structure with $\rho$=0.9 g/cm$^3$. As suggested in Marchis *et al.* (2006a), P-type asteroids could be dormant comets containing significant amount of water ice. In this case, the macro-porosity of (283) Emma could be significantly less than 50%. For instance, if the asteroid is composed of pure ice, its density will be less than 10% corresponding to a coherent internal structure. Spectroscopic studies, combining visible, near-infrared and far-infrared spectra could help to better estimate the surface composition and mineralogy of this P-type asteroid.

**4.3 Density of a C-type asteroid: (379) Huenna**

The orbit of Huenna's satellite is extremely well constrained in our study, since the measured positions are well distributed along the orbit (Fig. 3c). This asteroid is classified as C-type asteroid by Bus and Binzel (2002). Taking $D_{STM}$ = 92.3 km, we derived a low bulk density of 0.9±0.1 g/cm$^3$. Using the NEATM analysis ($D_{NEATM}$=97.6 km) its bulk density is even lower (0.8±0.1 g/cm$^3$). This result, based on 6 IRAS sightings, is consistent with our previously published C-type asteroid densities but it is also in agreement with the lower density of P-type asteroids. The discrepancy between the C-type asteroid bulk densities of (121) Hermione (Marchis *et al.* 2005b), (90) Antiope (Descamps *et al.* 2007) and the carbonaceous chondrite meteorites, assumed to be their meteorite analogs with a bulk density >2 g/cm$^3$, was interpreted by various authors as the result of a high macro-porosity (~30-60%). The (379) Huenna system, however, displays, conspicuous differences to those well-studied binary systems. Huenna's moonlet orbits far away from the primary in term of stability (~20% × $R_{Hill}$) and has a significant eccentricity (e~0.3) suggesting that the satellite is more likely a captured fragment. Therefore, the scenario of formation after disruption of a large parent asteroid and subsequent reaccretion of the primary may not apply in this case. However, significant macro-porosity measurements for minor planets have been reported on the basis of spacecraft observations, e.g. (253) Mathilde (Yeomans *et al.* 1997) and more recently (25143) Itokawa (Fujiwara *et al.* 2006). Because the presence of moonlet companions has not been reported for these asteroids, we can assume that a rubble-pile internal structure is not necessarily associated with a moonlet companion. (379) Huenna may have had a complex history. It could be the product of a disruption of a parent asteroid, which subsequently captured an interloper or remaining fragment of the parent disruption. This asteroid is a member of the Themis family (see Zappala et al. 1995). A spectroscopic comparison of the main asteroid and its satellite should help to constrain the origin of this system. Knowledge of its orbital elements facilitates optimization of the observations which can be performed only with an AO system mounted on a large telescope. For instance, we are now able to schedule them when the angular separation between the moonlet and the primary will be at its maximum.

**4.4 Bulk density of a S-type asteroid: 3749 Balam**

There is no radiometric measurement of (3749) Balam's effective diameter, neither by IRAS nor the Spitzer Space Telescopes. Since (3749) Balam is a member of the Flora collisional family (Zappalà *et al.* 1995) it is presumably an S-type asteroid. Assuming an albedo $p_v$ =0.15 and an H-value of 13.4, the corresponding equivalent diameter should be $D_{avg}$=7.2 km. Considering the average flux ratio in Table 5d and assuming the same albedo for the components, we can estimate their average diameters to be $D_{primary}$= 6.6±0.2 km and $D_{satellite}$= 2.8±0.4 km.

Using the average diameter we derived a bulk density of $\rho \sim 2.6$ g/cm$^3$ significantly higher than the densities of the main-belt multiple asteroidal systems studied so far and those presented here. This measurement is, however, in very good agreement with the bulk density of (433) Eros ($\rho$=2.67±0.03 g/cm$^3$, Wilkison *et al.* 2002), an S-type near-Earth asteroid intensively studied by the NEAR Shoemaker spacecraft. Taking OC meteorites as an analog with a bulk density of 3.4 g/cm3 and a microporosity between 0 and 15%, we derived a macro-porosity from 0 to 23% indicating that the (3749) Balam system is composed of coherent or poorly fractured components. This binary system is more likely the product of the mutual capture of two fragments produced by the disruption of proto-Flora asteroid, a 200-km diameter main-belt asteroid that disrupted ~1 billion years ago (Nesvorny *et al.* 2006).

## 5. Tidal effect dissipation

**Material**: include here Figure 4.

### 5.1 Orbital stability

Tidal dissipation between the satellite and the primary of a binary asteroid system can affect the orbital elements of the satellite. Based on previous work of Harris and Ward (1982), Weidenschilling *et al.* (1989) defined the condition of stability for a binary system if the two components have the same density:

$$\left(\frac{a}{R_p}\right)^2 < \frac{6}{5}\frac{(1+q)(1+q^{5/3})}{q} \quad (3)$$

where $q = M_s/M_p$ and $R_p$ is the radius of the primary. We do not have a direct measurement of $q$, but it can be estimated using the radius measurements ($q \sim (R_s/R_p)^3$ with $R_s$ the radius of the satellite), assuming the same bulk density for the two components of the system. From this equation, we conclude that the orbit of (3749) Balam is the only binary asteroid in which the companion is not perturbed by tidal dissipation effect. (379) Huenna is very close to being stable but we should allow for the possibility of different bulk density of the moon and the primary if the satellite is a captured interloper. We will therefore limit the study of tidal dissipation to (130) Elektra and (283) Emma. Figure 4 shows the domains of separation a/R vs mass ratio *q*. (130) Elektra and (283) Emma both fall well short of synchronous stability, indicating that the orbits of their moonlets will evolve due to tidal dissipation.

**5.2 Time scale for semi-major axes**

For a satellite that was formed outside the synchronous orbit ($a_{syn}$) the tides raised by the satellite on the primary will increase its semi-major axis (*a*) and decrease the spin rate of the primary ($\Omega$). From Kepler's law, we know that $a^3_{syn} = (GM_p/\Omega^2)$. So, if the spin of the primary slows down, the synchronous radius $a_{syn}$ will increase. The timescales for changes in *a* and $\Omega$ are not well constrained because the dissipation properties of an asteroid satellite are not well known. Weidenschilling *et al.* (1989) estimated the tidal evolution timescale $\tau$ from initial, $a_i$, to final semi-major axis, $a_f$, as

$$\left(\frac{a_f}{R_p}\right)^{13/2} - \left(\frac{a_i}{R_p}\right)^{13/2} = K\tau \frac{\rho^{5/2} q \sqrt{1+q} R_p^2}{\mu Q} \qquad (4)$$

where $K = 10\pi^{3/2} G^{3/2}$, $\rho$ is the bulk density, and $\mu Q$ is the tidal parameter, the product of rigidity ($\mu$) and specific dissipation parameter (Q). $\mu Q \sim 10^{10}$ is the best guess for this parameter product considering Q~100 as measured for Phobos by Yoder (1982) and $\mu$ ~$10^8$ N m$^{-2}$ a typical value for a moderately fractured asteroid. Durda *et al.* (2004) do not discuss the value of $a_i$ in their SPH simulations of collisions and formation of moonlet binary asteroids, but we can neglect the term $(a_i/R_p)^{13/2}$ since $(a_f/R_p) \sim 10$ from our analysis in the equation 4 and directly invert it to estimate the time scale $\tau$. We derive an approximate age for (130) Elektra and (283) Emma of greater than 4.5 billion years and ~10 million years respectively (see Fig. 4). Estimation of the age of these asteroids using for instance the modeling of their collisional family or reddening of the spectrum by space weathering is desirable since it could lead to the direct determination of μQ for a rubble pile asteroid. A large diversity of ages for collisional families have been already reported: 2.5 Byr for the large Themis family (Nesvorny *et al.* 2006) and a few hundred thousand years for the more recent ones (Nesvorny and Vokrouhlicky, 2006)

**5.3 Evolution of eccentricity**

Tidal evolution also modifies the satellite's eccentricity; the tidal forces on the satellite vary along the orbit and will tend to circularize the orbit, whereas the tide on the planet will increase the eccentricity. From Harris and Ward (1982), assuming that the physical properties (such as density, rigidity and Q) of the primary and secondary are similar, we derive:

$$\frac{\dot{e}}{e} = \left(\frac{19}{8}\text{sgn}(2\Omega - 3n) - \frac{7Rs}{2Rp}\right)\frac{\dot{a}}{a} \quad (5)$$

where *sgn* is the *sign* function. $\Omega$, the spin rate of the primary is derived from the orbital period ($P=2\pi\Omega$) measured accurately by lightcurve observations (see Table 1). Harris and Warner (Minor Planet Lightcurve Parameters[1]) report consistent measurements for (130) Elektra with P=5.22h from various sources. Although one measurement was published for (283) Emma in Stanzel (1978), it is clear that the synodic period of the primary spin is close to 6.88 h. In both cases, using the measured size ratio (Table 7), Eq. 5 indicates that the eccentricity will be excited and then increase. Both systems are located beneath the limit of *e* excitation in Fig. 4, therefore we can conclude that their observed eccentricities (~0.1) are most likely due to the tidal effect. Harris (1980) showed that in the case of a moonlet and a primary of the same composition, the rate of eccentricity growth depends on the semi-major axis and the eccentricity stalls at around ~0.7 at most. A modest

---

[1] http://cfa-www.harvard.edu/iau/lists/LightcurveDat.html

eccentricity of a few tenths seems realistic

**5.4 Application: rigidity coefficient of (283) Emma**

Nesvorny et al. (2005) identified (283) Emma as the largest member of a collisional family using a statistically-robust method. Emma collisional family is composed of 76 identified members and has a parent body with estimated diameter of 185 km. The precise age of the family could not have been determined because the family is located in a dynamically complicated region. Detailed analysis performed recently using modeling of family spreading via Yarkovsky thermal effect (Bottke et al., 2001) suggests that the approximate age of Emma family is ~300 Myr only. Using Eq. 4, we derive that $\mu Q$ ~$10^{11}$ is the best guess for Emma. Considering $Q$~100, then $\mu = 10^9$ N/m$^2$ = $10^{10}$ dynes/cm$^2$. This rigidity coefficient is close to the one for ice ($\mu_{Ice} = 2 \times 10^{10}$ dynes/cm$^2$ in Farinella et al., 1979). We estimated the bulk density of this P-type asteroid to be pretty low (0.9± 0.1 g/cm$^3$) similar to (617) Patroclus (Marchis et al., 2006). The rigidity coefficient calculated here suggests that (283) Emma could be also a dormant comet.

**6. Conclusions**

We have described the first orbit determination of four binary asteroidal systems located in the main-belt on the basis of adaptive optics observations collected with various 8-10m class telescopes. Their satellites clearly describe orbits with significant eccentricities. Because of the wide range of eccentricities observed in these systems (from 0.1 to 0.9), we propose different origin and evolution scenarios. Using the best-fitting orbital parameters, we have estimated the masses and the bulk densities of the systems:

- The (130) Elektra system is well characterized. Its companion S/2003(130)1 ($D_s$=7 km) orbits around the primary ($D_p$~200 km) at 1/40 × $R_{Hill}$ with a modest eccentricity of ~0.1 most likely due to excitation by the tidal effect. The satellite revolves around the primary in the opposite direction of the spin of the primary. The bulk density derived using the IRAS/STM diameter (~2 g/cm$^3$) is similar to that of (1) Ceres, another G-type asteroid. Because the primary is resolved in our AO data, we were able measure the bulk density; the result is a significantly lower value (~1.3 g/cm$^3$), but one that is in agreement with those reported for other binary C-type asteroids.

- (283) Emma is another binary system, whose companion (S/2001(283)1 with $D_s$~10 km) orbits close to the primary ($D_p$ ~ 140 km) with a modest eccentricity of 0.1. We also conclude that this system is evolving due to tides, and the eccentricity is due to excitation by the primary spin. The taxonomic class of (283) Emma is not well defined, but its albedo suggests that it should be a P-type asteroid. The bulk density (~0.9 g/cm$^3$) derived from the orbit analysis and the IRAS and AO diameters is of the same order than the bulk density of (617) Patroclus, another P-type asteroid, but located in the Trojan population. Considering the age of the Emma collisional family (~300 Myrs) we derive a coefficient of rigidity in agreement with an icy interior composition ($\mu = 10^{10}$ dynes/cm$^2$).

- The (379) Huenna binary system was very well constrained by our program. We derived a low bulk density (0.9-1.2 g/cm$^3$) indicative of a significant macro-porosity for this ~100-km C-type asteroid. However, the significant eccentricity (~0.3) suggests that the loosely bound satellite ($D_s$~6 km) is more likely a captured fragment.

- The (3749) Balam binary system is the only S-type asteroid in our study. The orbit of this loosely-bound binary system, which is composed of two components of roughly equal size, is not very well defined but should have a strong eccentricity (~0.9). Its bulk density (~2.6 g/cm$^3$) is very close to that measured for (433) Eros, another S-type asteroid visited by NEAR Shoemaker, indicating a coherent internal structure. This binary system is more likely formed by mutual capture of two coherent fragments after a large collision.

The orbits of these binary systems will be refined in the future with further observations provided by numerous AO systems now available on various 8-10m class telescopes. We expect to be able to extract low order perturbations, such as the precession of the orbit due to the irregular shape of the primary. Additionally, it may be possible to predict and observe mutual events between the components of a system which will help to estimate directly the size and shape of the primary; such work was performed by our team for (617) Patroclus-Menoetius (Marchis *et al.* 2007a). We might also expect to observe stellar occultations by the secondary, which would provide a direct measurement of its apparent diameter. Recent successful observations were reported by Soma *et al.* (2006) for Linus, satellite of (22) Kalliope.

In this work we pointed out the discrepancy between diameter estimate from IRAS measurements and AO observations. This has a significant impact on the calculated bulk density and the inferred porosity. Observations of these binary systems using FIR instruments (SPITZER telescope or the future SOFIA aircraft) combined with an accurate shape and pole model obtained by lightcurve inversion are keys to contrain these values. A better estimate of the size and shape of the primary and its satellite will help to establish the origin of the system, and to derive its bulk density and porosity, which are the two important physical parameters that can otherwise only be derived if the asteroid is visited by a spacecraft.

Since we have a good knowledge of the orbital parameters of various binary systems, we should be able to optimize spectroscopic observations of the primary and the moonlet using new integral field imagers or slit spectrograph combined with AO systems. A spectroscopic comparison will help to constrain the origin of the system knowing if the moonlet was captured and has a different composition than the primary or if it is a subsequent fragment of a large collision which also formed the primary.

**Acknowledgements**

We kindly thank the referees Petr Pravec and Schelte "Bobby" Bus for their constructive and accurate comments. This work was equally supported by the National Science Foundation Science and Technology Center for Adaptive Optics, and managed by the University of California at Santa Cruz under cooperative agreement No. AST-9876783 and by the national Aeronautics and Space Administration issue through the Science Mission Directorate Research and Analysis Programs number NNG05GF09G. Part of these data was obtained at the W.M. Keck observatory, which is operated as a scientific partnership between the California Institute of Technology, the University of California and the National Aeronautics and Space Administration. The observatory and its AO system were made possible by the generous financial support of the W. M. Keck Foundation. Other observations were obtained at the Gemini Observatory and the Gemini Science Archive, which is operated by the Association of Universities for Research in Astronomy, Inc., under a cooperative agreement with the NSF on behalf of the Gemini partnership. We are very thankful to Mikko Kaasalainen for his expertise in 3D-shape reconstruction and for providing Elektra shape model.

**References**

André, C.L.F., 1901, Sur le système formé par la Planète double (433) Eros , Astr. Nach., 155, 27

Babcock, H.W., 1953. The possibility of compensating seeing, PASP 65, 229-239

Behrend, R. , Bernasconi, L., Roy, R., and 46 collaborators,  2006. Four new binary minor planets: (854) Frostia, (1089) Tama, (1313) Berna, (4492) Debussy,  A&A, 446, 1177

Bottke, W.F., Vokrouhlicky, D. Broz, M., Nesvorny, D., & Morbidelli, A., 2001. Dynamical spreading of asteroid families by the Yarkovsky effect, Science 294, 5547, 1693-1696

Britt, D.T. & Consolmagno, G.J., 2003. Stony meteorite porosities and densities: A review of the data through 2001, MP&S, 38, 8, 1161-1180

Bus, S. J. and Binzel, R. P.,2002. Phase II of the Small Main-Belt Asteroid Spectroscopic Survey: A Feature-Based Taxonomy", Icarus 158, 146-177

Cuk. M. and Burns, J.A, 2005. Effects of thermal radiation on the dynamics of binary NEAs,  Icarus, 176, 2, 418-431

Descamps, P., Marchis, F., Berthier, J., Prangé, R., Fusco, T. and Le Guyader, C. 2002. First ground-based Astrometric Observations of Puck CRAS, Physique 3, p121-128, 2002.

Descamps, P. 2005. Orbit of an Astrometric Binary System. Celestial Mechanics and Dynamical Astronomy, 92, 381-402

Descamps, P., Marchis, F., Michalowski, T., *et al.* 2007. Figure of the double asteroid 90 Antiope from AO and lightcurves observations. Icarus, 189, 2, 362-369.

Devillard, N., 1997. The eclipse software, The Messenger 87 (1997), pp. 19–20.

Durda, D.D., Bottke Jr., W.F., Enke, B.L., Merline, W.J., Asphaug, E., Richardson, D.C., and Z.M. Leinhardt, 2004. The formation of asteroid satellites in large impacts: results from numerical simulations. Icarus 170, 243-257.

Durech, J. and 41 colaborators, 2007. Physical models of ten asteroids from observers collaboration network, A&A, 465, 1, 331-337

Farinella, P., Milani, A., Nobili, A.M., Valsecchi, G.B., 1979. Tidal Evolution and the Pluto-Charon system, Moon and the Planets, 20, 415-421

Harris, A.W & Ward, W.R. 1982. Dynamical constraints on the formation and evolution of planetary bodies, In: Annual review of earth and planetary sciences, 10., Palo Alto,

CA, Annual Reviews, Inc., 1982, 61-108.

Harris, A.W. 1998. A Thermal Model for Near-Earth Asteroids, Icarus 131, 291.

Harris, A.W. 2006. The surface properties of small asteroids from thermal-infrared observations, Asteroids, Comets, Meteors, Proceedings of the 229th Symposium of the International Astronomical Union held in Búzios, Rio de Janeiro, Brasil August 7-12, 2005, Edited by Daniela, L.; Sylvio Ferraz, M.; Angel, F. Julio Cambridge: Cambridge University Press, 449-463

Herriot, G., Morris, S., Anthony, A., Derdall, D., Duncan, D. Dunn, J., Ebbers, A.W., Fletcher, J.M., Hardy, T., Leckie, B., Mirza, A., Morbey, C.L., Pfleger, M., Roberts, S., Shott,, P., Smith, M., Saddlemyer, L.K., Sebesta, J., Szeto, K., Wooff, R., Windels, W., Veran, J.-P, 2000. Progress on ALTAIR: the Gemini North adaptive optics system, Proc. SPIE vol. 4007, 115-125, Adaptive Optical Systems Technology, Peter L. Wizinowich; Ed.

Hestroffer, D., Vachier, F., Balat, B. 2005. Orbit Determination of Binary Asteroids, EM&P. 97, 3-4, 245-260.

Hodapp , K.W., Jensen , J.B., Irwin , E.M., Yamada , H., Chung , R., Fletcher , K., Robertson , L., Hora , J.L., Simons , D.A., Mays , W., Nolan , R., Bec , M., Merrill , M., Fowler , A.M. 2003. The Gemini Near-Infrared Imager (NIRI), PASP 115, 1388-1406

Fujiwara,A., J. Kawaguchi, D. K. Yeomans, M. Abe, T. Mukai, T. Okada, J. Saito, H. Yano, M. Yoshikawa, D. J. Scheeres, O. Barnouin-Jha, A. F. Cheng, H. Demura, R. W. Gaskell, N. Hirata, H. Ikeda, T. Kominato, H. Miyamoto, A. M. Nakamura, R. Nakamura, S. Sasaki, K. Uesugi, 2006. The Rubble-Pile Asteroid Itokawa as Observed by Hayabusa, Science, 312, 5578, 1330-1334.

Hom, E. F. Y., F. Marchis, T. K. Lee, S. Haase, D. A. Agard, and J. W. Sedat 2007. An adaptive image deconvolution algorithm (AIDA) with application to multi-frame and 3d data. JOSA A, 24, 6, 1580-1600.

Kepler, J., 1609. Astronomia nova, Pragae.

Lazzaro, D., Angeli, C.A., Carvano, J.M., Mothe-Diniz, T., Duffard, R., Florczak, M. 2004. $S^3OS^2$: the visible spectroscopic survey of 820 asteroids, Icarus, Volume 172, Issue 1, p. 179-220

Lebofsky, L. A., Sykes, M. V., Tedesco, E. F., Veeder, G. J., Matson, D. L., Brown, R. H., Gradie, J. C., Feierberg, M. A., Rudy, R. J., 1986. A refined "standard" thermal model for asteroids based on observations of 1 Ceres and 2 Pallas. Icarus 68, 239 – 251.

Lenzen, R., Hartung, M., Brandner, W., Finger, G., Hubin, N.N., Lacombe, F., Lagrange, A.-M., Lehnert, M.D., Moorwood, A.F.M., Mouillet, D., 2003. instrument Design and Performance for Optical/Infrared Ground-based Telescopes. Edited by Iye M, Moorwood, A. F. M. Proceedings of the SPIE, Volume 4841, pp. 944-952

Marchis, F., Descamps, P., Hestroffer, D., Berthier, J., Vachier, F., Boccaletti, A., de Pater, I., Gavel, D., 2003, A three-dimensional solution for the orbit of the asteroidal satellite of 22 Kalliope. Icarus, 165, 112-120.

Marchis, F., Descamps, P., Hestroffer, D., Berthier, J., 2005a. Discovery of the triple asteroidal system 87 Sylvia. Nature, 436, 822-824.

Marchis, F., Hestroffer, Descamps, P., D., Berthier, J.,Laver, C., de Pater, I. 2005b. Mass and density of Asteroid 121 Hermione from an analysis of its companion orbit, Icarus, 178, 2, 450-464

Marchis, F., Hestroffer, D., Descamps, P., Berthier, J., Bouchez, A. H., Campbell, R. D., Chin, J. C. Y., van Dam, M. A., Hartman, S. K., Johansson, E. M., Lafon, R. E., Le Mignant, D., de Pater, Imke, Stomski, P. J., Summers, D. M., Vachier, F., Wizinovich, P. L., Wong, M. H., 2006a. A low density of 0.8 g.cm$^{-3}$ for the Trojan binary asteroid 617 Patroclus. Nature, 439, 565-567.

Marchis, F., Kaasalainen, M., Hom, E.F.Y., Berthier, J., Enriquez, J., Hestroffer, D., Le Mignant, D., and de Pater, I. 2006b. Shape, Size, and multiplicity of main-belt asteroids. I Keck Adaptive Optics Survey, Icarus, 185, 39-63.

Marchis, F., Baek, M., Berthier, J., Descamps, P., Hestroffer, D., Kaasalainen, M., Vachier, F., 2006c. Large adaptive optics of asteroids (LAOSA): Size, shape, and occasionally density via multiplicity. Workshop on Spacecraft Reconnaissance of Asteroid and Comet Interiors, Abstract #3042.

Marchis, F., Baek, M., Berthier, J., Descamps, P., Hestroffer, D., Vachier, F., Colas, F., Lecacheux, J., Reddy, V., Pino, F. 2007a, Central Bureau Electronic Telegrams, 836, 1, Ed. by Green, D. W. E.

Marchis, F., Baek, M., Descamps, P., Berthier, J., Hestroffer, D., Vachier, F., 2007b. S/2004(45)1, IAU Circ., 8817, 1, Edited by Green, D. W. E.

Margot, J.-L., M.C. Nolan, L.A.M. Benner, S.J. Ostro, R.F. Jurgens, J.D. Giorgini, M.A. Slade, D.B. Campbell, 2002. Binary Asteroid in the Near-Earth Object Population, Science 296, 5572, 1445-1448

Margot, J.-L. 2003. S/2003 (379) 1, IAU Circ. 8182, 1.

Merline, W.J., Close, L.M., Dumas, C., Chapman, C.R., Roddier, F., Menard, F., Slater, D.C., Duvert, G., Shelton, C., Morgan, T. 1999. Discovery of a moon orbiting the asteroid 45 Eugenia, Nature, 401, 565.

Merline, W.J., Close, L.M., Siegler, N. Dumas, C., Chapman, Rigaut, F., P.M., Terrell, D. and Menard F. Owen, W.M., and Slater, D.C. 2002a. S/2002 (3749) 1, IAU Circ., 7827, 1.

Merline, W.J., L.M. Close, Siegler, N., Dumas, C., Chapman, C.R., Rigaut, F., Menard, F., Owen, W.M., Slater, D.C., Durda, D.D., 2002b. Discovery of a Loosely-bound Companion to Main-belt Asteroid (3749) Balam, American Astronomical Society, DPS Meeting #34, #02.01; Bulletin of the American Astronomical Society, Vol. 34, p.835

Merline, W.J., Dumas, C., Siegler, N. Close, L.M., Chapman, C.R., Tamblyn, P.M., Terrell, D. and Menard F. 2003a. S/2003 (283) 1, IAU Circ., 8165, 1.

Merline, W.J., Tamblyn, P.M., Dumas, C., Close, L.M., Chapman, C.R., and Menard, F. 2003b. S/2003 (130) 1, IAU Circ., 8183, 1.

Nesvorny, D.; Vokrouhlicky, D. , 2006. New Candidates for Recent Asteroid Breakups, Astron. J., 132, 5, 1950-1958.

Nesvorny, D., Bottke, W. F., Vokrouhlicky, D., Morbidelli, A,, Jedicke, R., 2006. Asteroids, Comets, Meteors, Proceedings of the 229th Symposium of the International Astronomical Union held in Búzios, Rio de Janeiro, Brasil August 7-12, 2005, Edited by Daniela, L.; Sylvio Ferraz, M.; Angel, F. Julio Cambridge: Cambridge University Press, 289-299

Noll, K., 2006. Solar System binaries. Asteroids, Comets, Meteors, Proceedings of the 229th Symposium of the International Astronomical Union held in Búzios, Rio de Janeiro, Brasil Brazil August 7-12, 2005, Edited by Daniela, L., Sylvio Ferraz, M., Angel, F. Julio Cambridge: Cambridge University Press, 301-318

Pravec, P. & Harris, A. W., 2007. Binary Asteroid Population. 1. Angular Momentum Content, Icarus, 190, 1, 250-259

Richardson, D.C., and Walsh, K.J., 2006, Binary Minor Planets, Ann. Rev Planet. Sci., 34, 47-81.

Rousset, G., Lacombe, F., Puget, P., Hubin, N.N., Gendron, E., Fusco, T., Arsenault, R., Charton, J., Feautrier, P., Gigan, P., Kern, P.Y., Lagrange, A.-M., Madec, P.-Y., Mouillet, D., Rabaud, D., Rabou, P., Stadler, E., Zins, G. 2003. Adaptive Optical System Technologies II. Edited by Wizinowich, Peter L.; Bonaccini, Domenico. Proceedings of the SPIE, Volume 4839, pp. 140-149

Stanzel, R. 1978. Lightcurve and Rotation Period of Minor Planet 283 Emma, Astron. Astrophys. Suppl., 34, 373-376.

Soma, M., Hayamizu, T., Berthier, J., Lecacheux, J. 2006. (22) Kalliope and (22) Kalliope I, Central Bureau Electronic Telegrams, 732, 1. Edited by Green, D. W. E.

Thomas, P.C., Parker, J.Wm, McFadden, L.A., Russell, C.T., Stern, S.A., Sykes, M.V., Young, E.F., 2005. Differentiation of the asteroid Ceres as revealed by its shape, Nature, 437, 7056, 224-226.

Tholen, D.J and Barucci, M.A., 1989. Asteroid taxonomy, In Asteroids II (R.P. Binzel, *et al.* eds), pp. 806-825. Univ. of Arizona, Tucson.

Tedesco, E.F., Noah, P.V., Noah, M. and Price, S.D. 2002. The supplemental IRAS Minor Planet Survey, Astron. J. 123, 1056-1085.

Van Flandern, T.C. *et al.* 1979. Satellites of asteroids, in Asteroids, Univ. of Arizona Press, 443-465

Yeomans, D.K., J.P., Barriot, D.W. Dunham, R. W. Farquhar, J. D. Giorgini, C. E. Helfrich, A. S. Konopliv, J. V. McAdams, J. K. Miller, W. M. Owen Jr., D. J. Scheeres, S. P. Synnott, B. G. Williams, 1997. Estimating the Mass of Asteroid 253 Mathilde from Tracking Data During the NEAR Flyby, Science, 278, 5346, 2106-2109.

Yoder, C.F., 1982. Tidal rigidity of PHOBOS, Icarus, 49, 327-346

Veillet, C., Parker, J.Wm, Griffin, I., Marsden, B., Doressoundiram, A., Buie, M., Tholen, D.J., Connelley, M., Holman, M.J., 2002. The binary Kuiper-belt object 1998 WW31, Nature, 416, 6882, 711-713

Weidenschilling, S.J., Paolicchi, P.Zappala, V. 1989. Do asteroids have satellites? In Asteroids II; Proceedings of the Conference, Tucson, AZ, Mar. 8-11, 1988 (A90-27001 10-91). Tucson, AZ, University of Arizona Press, 1989, p. 643-658.

Walsh, K.J. & Richardson, D.C., 2006. Steady-state Population Of The Nea Binaries And Yorp Spinup Models, AAS-DPS 38, #53.08

Wilkison, S.L., Robinson, M.S., Thomas, P.C., Veverka, J., McCoy, T.J., Murchie, S.L., Prokter, L.M., Yeomans, D.K. 2002. An Estimate of Eros's Porosity and Implications for Internal Structure, Icarus, Volume 155, 94-103

Zappalà, V., Bendjoya, Ph., Cellino A., Farinella P., Froeschlé C., 1995. Asteroid families: Search of a 12,487-asteroid sample using two different clustering techniques, Icarus 116, 291-314.

**Table 1**
Characteristics of the studied binary minor planets. IRAS radiometric diameters (and their 1-σ uncertainty) are estimated using STM or NEATM models.

| Asteroid | Primary Diameter (km) | | | Rotational Period / max(a/b)$^5$ hours | Sp. type | Secondary | |
|---|---|---|---|---|---|---|---|
| | IRAS STM | IRAS NEATM | AO | | | Name | $D_{satellite}$ |
| 130 Elektra | 182±12 | 196±11 | 215±15 | 5.225/1.58 | G$^1$ | S/2003 (130)1 | 7±3 |
| 283 Emma | 148±5 | 141±6 | 160±10 | 6.888/1.31 | X$^1$ | S/2003 (283)1 | 9±5 |
| 379 Huenna | 92±2 | 98±3 | n/a | 7.002/1.09 | C$^2$ | S/2003 (379)1 | 5.8±1.2 |
| 3749 Balam | n/a | n/a | n/a | unk. | S$^3$ | S/2002 (3749)1 | 5.2±1 |

1. Tholen and Barucci, (1989)
2. Bus and Binzel, (2002)
3. Member of the Flora family
4. Tedesco *et al.* (2002)
5. Minor Planet Lightcurve Parameters, A.W. Harris and B. D. Warner, http://cfa-www.harvard.edu/iau/lists/LightcurveDat.html

**Table 2a**

Summary of our AO Observations of (130) Elektra and (283) Emma collected with the Keck, VLT, or Gemini North telescopes. The predicted magnitude in visible ($m_v$), celestial coordinates (RA, DEC), and distance from Earth are extracted from the IMCCE ephemeris web site (http://www.imcce.fr).

| ID | Name | Date | UT | Telescope | Filter | mv predicted | Airmass | RA | DEC | Distance from Earth (AU) |
|---|---|---|---|---|---|---|---|---|---|---|
| 130 | Elektra | 07-Dec-03 | 07:16:10 | Keck | Kp | 11.2 | 1.43 | 03 45 25.03 | -15 58 16.9 | 1.73829 |
| 130 | Elektra | 05-Jan-04 | 02:59:13 | VLT | Ks | 11.5 | 1.12 | 03 34 37.62 | -11 48 44.5 | 1.97767 |
| 130 | Elektra | 05-Jan-04 | 04:25:39 | VLT | Ks | 11.5 | 1.42 | 03 34 37.44 | -11 48 03.1 | 1.97831 |
| 130 | Elektra | 06-Jan-04 | 03:06:56 | VLT | Ks | 11.7 | 1.14 | 03 34 36.68 | -11 37 09.7 | 1.98832 |
| 130 | Elektra | 07-Jan-04 | 04:53:27 | VLT | Ks | 11.7 | 1.68 | 03 34 37.27 | -11 24 39.1 | 1.99984 |
| 130 | Elektra | 07-Jan-04 | 05:05:34 | VLT | Ks | 11.7 | 1.79 | 03 34 37.27 | -11 24 33.2 | 1.99994 |
| 130 | Elektra | 07-Jan-04 | 05:13:04 | VLT | Ks | 11.7 | 1.88 | 03 34 37.28 | -11 24 29.5 | 1.99999 |
| 130 | Elektra | 07-Feb-04 | 07:09:00 | Keck | Kp | 12.1 | 1.28 | 03 46 34.70 | -04 52 43.1 | 2.38090 |
| 130 | Elektra | 02-Mar-04 | 00:26:40 | VLT | H | 12.4 | 1.33 | 04 08 00.59 | -00 01 38.5 | 2.70366 |
| 130 | Elektra | 02-Mar-04 | 00:30:54 | VLT | H | 12.4 | 1.34 | 04 08 00.77 | -00 01 36.4 | 2.70370 |
| 130 | Elektra | 30-Oct-04 | 15:03:40 | Gemini | Kp | 13.2 | 1.28 | 09 48 53.38 | 06 38 53.9 | 3.38516 |
| 130 | Elektra | 30-Oct-04 | 15:05:59 | Gemini | Kp | 13.2 | 1.27 | 09 48 53.47 | 06 38 53.7 | 3.38514 |
| 130 | Elektra | 30-Oct-04 | 15:10:16 | Gemini | Kp | 13.2 | 1.26 | 09 48 53.62 | 06 38 53.2 | 3.38510 |
| 130 | Elektra | 02-Nov-04 | 15:28:32 | Gemini | Kp | 13.2 | 1.17 | 09 51 30.72 | 06 31 19.0 | 3.34907 |
| 130 | Elektra | 02-Nov-04 | 15:34:18 | Gemini | Kp | 13.2 | 1.16 | 09 51 30.92 | 06 31 18.4 | 3.34902 |
| 130 | Elektra | 03-Nov-04 | 15:33:28 | Gemini | Kp | 13.2 | 1.15 | 09 52 21.59 | 06 28 55.4 | 3.33692 |
| 130 | Elektra | 05-Nov-04 | 15:24:06 | Gemini | Kp | 13.2 | 1.16 | 09 54 00.42 | 06 24 22.1 | 3.31259 |
| 130 | Elektra | 15-Jan-05 | 12:25:31 | Keck | Kp | 12.5 | 1.03 | 10 09 43.96 | 08 39 52.5 | 2.49434 |
| 130 | Elektra | 15-Jan-05 | 14:14:01 | Keck | Kp | 12.5 | 1.08 | 10 09 41.80 | 08 40 24.5 | 2.49380 |
| 130 | Elektra | 12-Mar-06 | 13:38:32 | Gemini | Kp | 13 | 1.02 | 13 53 02.97 | 12 35 54.7 | 2.93475 |
| 130 | Elektra | 08-Apr-06 | 12:03:05 | Gemini | Kp | 12.8 | 1.04 | 13 37 55.90 | 16 10 06.0 | 2.84796 |
| 130 | Elektra | 08-Apr-06 | 12:08:40 | Gemini | Kp | 12.8 | 1.05 | 13 37 55.74 | 16 10 07.5 | 2.84796 |
| 130 | Elektra | 09-Apr-06 | 09:12:20 | Gemini | Kp | 12.8 | 1.09 | 13 37 20.23 | 16 15 49.1 | 2.84854 |
| 130 | Elektra | 09-Apr-06 | 09:20:23 | Gemini | Kp | 12.8 | 1.08 | 13 37 20.00 | 16 15 51.2 | 2.84854 |
| 130 | Elektra | 11-Apr-06 | 06:01:03 | VLT | Ks | 12.8 | 1.37 | 13 36 03.80 | 16 27 34.6 | 2.85048 |
| 130 | Elektra | 11-Apr-06 | 06:11:58 | VLT | H | 12.8 | 1.39 | 13 36 03.48 | 16 27 37.3 | 2.85049 |
| 130 | Elektra | 11-Apr-06 | 06:22:16 | VLT | J | 12.8 | 1.42 | 13 36 03.18 | 16 27 40.0 | 2.85050 |
| 130 | Elektra | 11-Apr-06 | 11:08:12 | Gemini | Kp | 12.8 | 1.01 | 13 35 55.08 | 16 28 50.6 | 2.85075 |
| 130 | Elektra | 12-Apr-06 | 11:56:27 | Gemini | Kp | 12.8 | 1.06 | 13 35 12.60 | 16 35 03.5 | 2.85230 |
| 130 | Elektra | 13-Apr-06 | 12:16:56 | Gemini | Kp | 12.8 | 1.10 | 13 34 30.86 | 16 40 59.3 | 2.85411 |
| 130 | Elektra | 27-Apr-06 | 03:33:18 | VLT | Ks | 12.9 | 1.36 | 13 25 20.96 | 17 43 00.3 | 2.90513 |
| 130 | Elektra | 27-Apr-06 | 06:20:43 | Gemini | Kp | 12.9 | 1.42 | 13 25 16.62 | 17 43 20.5 | 2.90578 |
| 130 | Elektra | 28-Apr-06 | 03:39:12 | VLT | Ks | 12.9 | 1.35 | 13 24 42.58 | 17 46 10.8 | 2.91078 |
| 130 | Elektra | 30-Apr-06 | 03:26:54 | VLT | Ks | 12.9 | 1.36 | 13 23 27.90 | 17 51 53.4 | 2.92270 |
| 130 | Elektra | 02-May-06 | 04:48:53 | VLT | Ks | 12.9 | 1.46 | 13 22 12.90 | 17 56 58.1 | 2.93600 |
| 130 | Elektra | 15-May-06 | 09:49:07 | Gemini | Kp | 13.1 | 1.09 | 13 15 19.81 | 18 10 05.8 | 3.04298 |
| 130 | Elektra | 16-May-06 | 08:14:44 | Gemini | Kp | 13.1 | 1.00 | 13 14 55.85 | 18 09 48.1 | 3.05182 |
| 130 | Elektra | 20-May-06 | 01:58:18 | VLT | Ks | 13.1 | 1.36 | 13 13 27.43 | 18 07 07.0 | 3.08873 |
| 130 | Elektra | 20-May-06 | 02:08:18 | VLT | H | 13.1 | 1.36 | 13 13 27.26 | 18 07 06.6 | 3.08880 |
| 130 | Elektra | 20-May-06 | 02:18:18 | VLT | J | 13.1 | 1.36 | 13 13 27.10 | 18 07 06.1 | 3.08887 |
| 130 | Elektra | 23-May-06 | 03:04:13 | VLT | Ks | 13.1 | 1.43 | 13 12 24.74 | 18 03 08.3 | 3.12046 |
| 130 | Elektra | 28-May-06 | 01:46:51 | VLT | Ks | 13.3 | 1.36 | 13 11 01.98 | 17 53 28.5 | 3.17485 |
| 130 | Elektra | 29-May-06 | 02:30:30 | VLT | Ks | 13.3 | 1.41 | 13 10 47.71 | 17 50 59.1 | 3.18660 |
| 130 | Elektra | 02-Jun-06 | 02:03:32 | VLT | Ks | 13.3 | 1.39 | 13 10 02.67 | 17 39 54.5 | 3.23314 |
| 283 | Emma | 15-Jul-03 | 06:55:27 | VLT | H | 12.9 | 1.02 | 21 23 45.13 | -14 13 16.4 | 1.75561 |
| 283 | Emma | 15-Jul-03 | 07:13:31 | VLT | H | 12.9 | 1.03 | 21 23 44.64 | -14 13 16.0 | 1.75552 |
| 283 | Emma | 15-Jul-03 | 07:17:02 | VLT | Ks | 12.9 | 1.03 | 21 23 44.55 | -14 13 16.0 | 1.75551 |
| 283 | Emma | 15-Jul-03 | 07:20:20 | VLT | J | 12.9 | 1.03 | 21 23 44.46 | -14 13 15.9 | 1.75549 |
| 283 | Emma | 15-Jul-03 | 10:12:30 | VLT | H | 12.9 | 1.63 | 21 23 39.86 | -14 13 12.2 | 1.75469 |
| 283 | Emma | 15-Jul-03 | 10:13:31 | VLT | H | 12.9 | 1.64 | 21 23 39.83 | -14 13 12.2 | 1.75469 |
| 283 | Emma | 16-Jul-03 | 10:02:43 | VLT | Ks | 12.9 | 1.54 | 21 23 02.63 | -14 12 47.6 | 1.74807 |
| 283 | Emma | 16-Jul-03 | 10:27:27 | VLT | Ks | 12.9 | 1.77 | 21 23 01.95 | -14 12 47.2 | 1.74796 |
| 283 | Emma | 30-Oct-04 | 12:16:22 | Gemini | Kp | 13.4 | 1.04 | 05 05 55.32 | 32 40 38.9 | 1.98151 |
| 283 | Emma | 30-Oct-04 | 12:20:55 | Gemini | Kp | 13.4 | 1.03 | 05 05 55.23 | 32 40 39.0 | 1.98149 |
| 283 | Emma | 30-Oct-04 | 14:03:46 | Gemini | Kp | 13.4 | 1.07 | 05 05 53.11 | 32 40 41.6 | 1.98097 |
| 283 | Emma | 30-Oct-04 | 15:23:30 | Gemini | Kp | 13.4 | 1.24 | 05 05 51.48 | 32 40 43.2 | 1.98057 |
| 283 | Emma | 02-Nov-04 | 15:20:11 | Gemini | Kp | 13.2 | 1.27 | 05 04 20.70 | 32 42 06.6 | 1.95973 |
| 283 | Emma | 05-Nov-04 | 10:30:55 | Gemini | Kp | 13.2 | 1.14 | 05 02 42.88 | 32 42 34.0 | 1.94192 |
| 283 | Emma | 14-Nov-04 | 06:31:31 | VLT | Ks | 13 | 1.86 | 04 56 16.05 | 32 38 01.8 | 1.89735 |
| 283 | Emma | 15-Nov-04 | 05:42:46 | VLT | Ks | 13 | 1.85 | 04 55 27.66 | 32 36 55.8 | 1.89363 |
| 283 | Emma | 16-Nov-04 | 04:58:46 | VLT | Ks | 13 | 1.93 | 04 54 38.07 | 32 35 42.1 | 1.89014 |
| 283 | Emma | 16-Nov-04 | 05:56:37 | VLT | Ks | 13 | 1.84 | 04 54 35.90 | 32 35 38.9 | 1.89000 |
| 283 | Emma | 17-Nov-04 | 05:08:30 | VLT | Ks | 13 | 1.89 | 04 53 45.45 | 32 34 17.7 | 1.88677 |
| 283 | Emma | 18-Nov-04 | 06:19:18 | VLT | Ks | 13 | 1.87 | 04 52 49.47 | 32 32 41.0 | 1.88355 |
| 283 | Emma | 07-Dec-04 | 03:38:55 | VLT | Ks | 12.7 | 1.84 | 04 34 30.58 | 31 39 00.3 | 1.87780 |
| 283 | Emma | 07-Dec-04 | 03:55:49 | VLT | H | 12.7 | 1.81 | 04 34 29.87 | 31 38 57.5 | 1.87783 |
| 283 | Emma | 07-Dec-04 | 04:11:39 | VLT | J | 12.7 | 1.80 | 04 34 29.21 | 31 38 54.9 | 1.87785 |
| 283 | Emma | 08-Dec-04 | 04:17:35 | VLT | Ks | 12.7 | 1.80 | 04 33 30.83 | 31 34 52.4 | 1.88042 |
| 283 | Emma | 10-Dec-04 | 05:46:52 | VLT | Ks | 12.7 | 2.10 | 04 31 32.55 | 31 26 16.2 | 1.88663 |
| 283 | Emma | 14-Dec-04 | 03:55:29 | VLT | Ks | 12.8 | 1.78 | 04 27 56.19 | 31 08 56.1 | 1.90182 |
| 283 | Emma | 14-Dec-04 | 04:12:42 | VLT | H | 12.8 | 1.79 | 04 27 55.53 | 31 08 52.8 | 1.90187 |
| 283 | Emma | 14-Dec-04 | 04:32:18 | VLT | J | 12.8 | 1.83 | 04 27 54.77 | 31 08 49.0 | 1.90193 |
| 283 | Emma | 19-Dec-04 | 01:50:48 | VLT | Ks | 12.8 | 1.26 | 04 23 45.63 | 30 45 49.0 | 1.92700 |
| 283 | Emma | 19-Dec-04 | 03:29:22 | VLT | Ks | 12.8 | 1.76 | 04 23 42.19 | 30 45 29.3 | 1.92739 |
| 283 | Emma | 20-Dec-04 | 01:42:25 | VLT | Ks | 12.8 | 1.96 | 04 22 58.22 | 30 41 00.2 | 1.93290 |
| 283 | Emma | 20-Dec-04 | 04:32:17 | VLT | Ks | 12.8 | 1.91 | 04 22 52.44 | 30 40 25.8 | 1.93362 |
| 283 | Emma | 28-Dec-04 | 02:44:43 | VLT | Ks | 13 | 1.73 | 04 17 22.87 | 30 01 17.9 | 1.99017 |
| 283 | Emma | 28-Dec-04 | 04:52:39 | VLT | Ks | 13 | 2.26 | 04 17 19.53 | 30 00 51.1 | 1.99090 |
| 283 | Emma | 26-Apr-06 | 06:07:19 | Gemini | Kp | 14.7 | 1.03 | 09 25 58.89 | 09 58 26.6 | 3.06803 |
| 283 | Emma | 18-May-06 | 06:02:06 | Gemini | Kp | 14.9 | 1.13 | 09 35 42.28 | 09 15 59.8 | 3.39509 |
| 283 | Emma | 07-Jun-06 | 05:58:31 | Gemini | Kp | 15.1 | 1.34 | 09 49 58.75 | 08 06 44.3 | 3.68627 |
| 283 | Emma | 11-Jun-06 | 06:10:17 | Gemini | Kp | 15.1 | 1.48 | 09 53 19.09 | 07 49 38.8 | 3.74194 |

**Table 2b:**
Summary of our AO Observations of (379) Huenna and (3749) Balam collected with the VLT-UT4 (Yepun) telescope and its NACO instrument.

| ID | Name | Date | UT | Telescope | Filter | mv predicted | Airmass | RA | DEC | Distance from Earth (AU) |
|---|---|---|---|---|---|---|---|---|---|---|
| 379 | Huenna | 08-Dec-04 | 07:08:41 | VLT | Ks | 14.2 | 1.38 | 07 47 38.48 | 19 01 48.1 | 2.42220 |
| 379 | Huenna | 09-Dec-04 | 06:35:44 | VLT | Ks | 14.2 | 1.41 | 07 47 08.91 | 19 02 46.9 | 2.41509 |
| 379 | Huenna | 09-Dec-04 | 06:48:16 | VLT | Ks | 14.2 | 1.40 | 07 47 08.63 | 19 02 47.5 | 2.41502 |
| 379 | Huenna | 10-Dec-04 | 06:51:34 | VLT | Ks | 14.2 | 1.39 | 07 46 36.99 | 19 03 51.4 | 2.40796 |
| 379 | Huenna | 14-Dec-04 | 05:28:48 | VLT | Ks | 14.0 | 1.51 | 07 44 20.77 | 19 08 36.1 | 2.38245 |
| 379 | Huenna | 14-Dec-04 | 07:09:01 | VLT | Ks | 14.0 | 1.39 | 07 44 18.10 | 19 08 41.7 | 2.38203 |
| 379 | Huenna | 15-Dec-04 | 05:20:30 | VLT | Ks | 14.0 | 1.53 | 07 43 43.55 | 19 09 55.9 | 2.37662 |
| 379 | Huenna | 28-Dec-04 | 05:37:03 | VLT | Ks | 13.8 | 1.40 | 07 34 09.23 | 19 31 11.3 | 2.32458 |
| 379 | Huenna | 28-Dec-04 | 07:41:22 | VLT | Ks | 13.8 | 1.61 | 07 34 04.85 | 19 31 20.8 | 2.32440 |
| 379 | Huenna | 29-Dec-04 | 05:13:41 | VLT | Ks | 13.8 | 1.41 | 07 33 20.60 | 19 33 00.01 | 2.32259 |
| 379 | Huenna | 18-Jan-05 | 03:58:39 | VLT | Ks | 13.4 | 1.41 | 07 16 08.30 | 20 11 48.2 | 2.34581 |
| 379 | Huenna | 18-Jan-05 | 06:17:38 | VLT | Ks | 13.4 | 1.71 | 07 16 03.31 | 20 11 58.9 | 2.34622 |
| 379 | Huenna | 21-Jan-05 | 02:25:32 | VLT | Ks | 13.7 | 1.56 | 07 13 43.41 | 20 17 16.6 | 2.35952 |
| 379 | Huenna | 25-Jan-05 | 04:51:45 | VLT | Ks | 13.7 | 1.51 | 07 10 30.17 | 20 24 37.7 | 2.38292 |
| 379 | Huenna | 25-Jan-05 | 06:43:58 | VLT | Ks | 13.7 | 2.26 | 07 10 26.53 | 20 24 45.5 | 2.38342 |
| 379 | Huenna | 26-Jan-05 | 02:47:49 | VLT | Ks | 13.7 | 1.45 | 07 09 49.04 | 20 26 12.8 | 2.38879 |
| 379 | Huenna | 26-Jan-05 | 05:10:53 | VLT | Ks | 13.7 | 1.58 | 07 09 44.45 | 20 26 23.0 | 2.38945 |
| 379 | Huenna | 27-Jan-05 | 03:10:56 | VLT | Ks | 13.7 | 1.42 | 07 09 03.97 | 20 27 57.1 | 2.39560 |
| 379 | Huenna | 27-Jan-05 | 06:08:04 | VLT | Ks | 13.7 | 1.97 | 07 08 58.41 | 20 28 09.4 | 2.39645 |
| 379 | Huenna | 28-Jan-05 | 03:04:48 | VLT | Ks | 13.7 | 1.42 | 07 08 20.73 | 20 29 37.8 | 2.40255 |
| 379 | Huenna | 28-Jan-05 | 03:14:05 | VLT | H | 13.7 | 1.42 | 07 08 20.44 | 20 29 38.4 | 2.40260 |
| 379 | Huenna | 28-Jan-05 | 03:22:34 | VLT | J | 13.7 | 1.42 | 07 08 20.18 | 20 29 39.0 | 2.40264 |
| 379 | Huenna | 02-Feb-05 | 03:09:22 | VLT | Ks | 13.9 | 1.42 | 07 04 57.90 | 20 37 39.4 | 2.44155 |
| 379 | Huenna | 02-Feb-05 | 05:09:40 | VLT | Ks | 13.9 | 1.74 | 07 04 54.62 | 20 37 46.9 | 2.44226 |
| 379 | Huenna | 04-Feb-05 | 02:41:11 | VLT | Ks | 13.9 | 1.42 | 07 03 44.92 | 20 40 38.5 | 2.45880 |
| 379 | Huenna | 04-Feb-05 | 04:06:03 | VLT | Ks | 13.9 | 1.52 | 07 03 42.73 | 20 40 43.7 | 2.45933 |
| 379 | Huenna | 04-Feb-05 | 04:14:50 | VLT | H | 13.9 | 1.54 | 07 03 42.50 | 20 40 44.2 | 2.45939 |
| 379 | Huenna | 04-Feb-05 | 04:23:59 | VLT | J | 13.9 | 1.57 | 07 03 42.27 | 20 40 44.7 | 2.45945 |
| 379 | Huenna | 07-Feb-05 | 03:44:25 | VLT | Ks | 13.9 | 1.50 | 07 02 01.44 | 20 45 00.3 | 2.48725 |
| 379 | Huenna | 08-Feb-05 | 02:30:20 | VLT | Ks | 13.9 | 1.42 | 07 01 31.56 | 20 46 18.6 | 2.49657 |
| 379 | Huenna | 08-Feb-05 | 02:45:38 | VLT | Ks | 13.9 | 1.42 | 07 01 31.22 | 20 46 19.5 | 2.49668 |
| 379 | Huenna | 09-Feb-05 | 03:16:53 | VLT | Ks | 14.1 | 1.46 | 07 01 00.19 | 20 47 41.9 | 2.50696 |
| 379 | Huenna | 16-Feb-05 | 01:21:14 | VLT | Ks | 14.1 | 1.45 | 06 58 06.00 | 20 56 09.7 | 2.58270 |
| 3749 | Balam | 15-Jul-03 | 05:30:15 | VLT | Ks | 16.5 | 1.09 | 18 32 27.84 | -250909.6 | 1.48401 |
| 3749 | Balam | 16-Jul-03 | 04:22:13 | VLT | Ks | 16.5 | 1.01 | 18 32 28.22 | -25 07 35.7 | 1.48736 |
| 3749 | Balam | 14-Nov-04 | 06:03:30 | VLT | Ks | 15.7 | 1.76 | 03 43 26.84 | 28 47 13.2 | 1.09647 |
| 3749 | Balam | 15-Nov-04 | 03:38:09 | VLT | Ks | 15.7 | 1.82 | 03 42 26.34 | 28 43 11.5 | 1.09440 |
| 3749 | Balam | 15-Nov-04 | 04:05:35 | VLT | H | 15.7 | 1.73 | 03 42 24.98 | 28 43 06.4 | 1.09436 |
| 3749 | Balam | 16-Nov-04 | 05:30:58 | VLT | Ks | 15.7 | 1.70 | 03 41 12.59 | 28 38 08.8 | 1.09218 |
| 3749 | Balam | 17-Nov-04 | 04:39:31 | VLT | Ks | 15.7 | 1.67 | 03 40 06.66 | 28 33 26.5 | 1.09045 |
| 3749 | Balam | 22-Nov-04 | 03:09:25 | VLT | Ks | 15.5 | 1.76 | 03 34 27.72 | 28 06 41.8 | 1.08529 |
| 3749 | Balam | 02-Dec-04 | 03:50:18 | VLT | Ks | 15.7 | 1.62 | 03 23 39.62 | 27 01 55.2 | 1.09370 |
| 3749 | Balam | 03-Dec-04 | 04:02:53 | VLT | Ks | 15.7 | 1.63 | 03 22 40.59 | 26 54 54.2 | 1.09592 |
| 3749 | Balam | 07-Dec-04 | 03:02:08 | VLT | Ks | 15.7 | 1.59 | 03 19 04.78 | 26 27 00.9 | 1.10697 |
| 3749 | Balam | 09-Dec-04 | 03:32:49 | VLT | Ks | 15.7 | 1.61 | 03 17 25.09 | 26 12 42.1 | 1.11402 |
| 3749 | Balam | 10-Dec-04 | 02:44:19 | VLT | Ks | 15.7 | 1.58 | 03 16 40.38 | 26 05 52.8 | 1.11772 |
| 3749 | Balam | 14-Dec-04 | 02:48:59 | VLT | Ks | 15.9 | 1.57 | 03 13 55.38 | 25 37 59.8 | 1.13521 |
| 3749 | Balam | 14-Dec-04 | 03:25:21 | VLT | Ks | 15.9 | 1.61 | 03 13 54.37 | 25 37 49.4 | 1.13533 |
| 3749 | Balam | 20-Dec-04 | 01:12:00 | VLT | Ks | 15.9 | 1.60 | 03 10 57.49 | 24 58 50.3 | 1.16716 |

**Table 3a**: Search for moonlet companions around (130) Elektra and (283) Emma. The characteristics of the 2-σ detection curve for each asteroid are calculated. α is the slope of the function, and $r_{lim}$ separation between both noise regimes dominated by the Poisson noise close to the primary at $r < r_{lim}$ and by the [detector+sky] noises at $r > r_{lim}$ At $r > r_{lim}$ the detection function can be approximated by a flat function with a value of $\Delta m_{lim}$ The radius of the Hill sphere is calculate based on consideration about the diameter and density of the asteroid (see Table 7 and details in Marchis *et al.* 2006b). The minimum diameter size for a moonlet to be detected at 1/4 and 2/100 $R_{Hill}$ is also indicated.

| ID | Name | Date | UT | α | Δmlim | r lim arcsec | Int Time s | Airmass | FWHM arcsec | Δm at 2/100xRHill | Diameter at 2/100XRHill | Δm at 1/4xRHill | Diameter at 1/4xRHill |
|---|---|---|---|---|---|---|---|---|---|---|---|---|---|
| 130 | Elektra | 07-Dec-03 | 07:16:10 | -4.0 | -8.5 | 0.87 | 540 | 1.43 | 0.16 | -7.7 | 5.2 | -8.5 | 3.6 |
| 130 | Elektra | 05-Jan-04 | 02:59:13 | -3.3 | -8.7 | 1.01 | 240 | 1.12 | 0.13 | -7.9 | 4.8 | -8.7 | 3.3 |
| 130 | Elektra | 05-Jan-04 | 04:25:39 | -3.7 | -8.5 | 0.94 | 240 | 1.42 | 0.14 | -7.5 | 5.8 | -8.5 | 3.7 |
| 130 | Elektra | 06-Jan-04 | 03:06:56 | -3.5 | -8.6 | 1.03 | 240 | 1.14 | 0.13 | -7.6 | 5.6 | -8.6 | 3.5 |
| 130 | Elektra | 07-Jan-04 | 04:53:27 | -4.3 | -7.3 | 0.97 | 240 | 1.67 | 0.27 | -5.7 | 13.2 | -7.4 | 6.2 |
| 130 | Elektra | 07-Jan-04 | 05:05:34 | -3.2 | -7.1 | 1.09 | 240 | 1.79 | 0.18 | -5.1 | 17.1 | -7.2 | 6.6 |
| 130 | Elektra | 07-Jan-04 | 05:13:04 | -2.9 | -6.8 | 1.19 | 240 | 1.88 | 0.19 | -5.1 | 17.6 | -6.8 | 8.1 |
| 130 | Elektra | 02-Mar-04 | 00:30:54 | -4.6 | -7.6 | 1.15 | 240 | 1.34 | 0.23 | -5.4 | 15.5 | -7.6 | 5.4 |
| 130 | Elektra | 30-Oct-04 | 15:03:40 | -17.4 | -5.9 | 0.35 | 40 | 1.28 | 0.16 | -5.7 | 13.5 | -6.8 | 8.0 |
| 130 | Elektra | 02-Nov-04 | 15:28:32 | -6.6 | -9.0 | 0.79 | 120 | 1.17 | 0.15 | -6.2 | 10.4 | -9.0 | 2.9 |
| 130 | Elektra | 03-Nov-04 | 15:33:28 | -5.6 | -9.2 | 1.05 | 120 | 1.15 | 0.13 | -6.3 | 10.2 | -9.2 | 2.7 |
| 130 | Elektra | 05-Nov-04 | 15:24:06 | -8.0 | -9.2 | 0.72 | 150 | 1.16 | 0.14 | -6.1 | 11.1 | -9.1 | 2.7 |
| 130 | Elektra | 15-Jan-05 | 12:25:31 | -8.2 | -8.6 | 0.72 | 180 | 1.03 | 0.12 | -7.2 | 6.8 | -8.7 | 3.3 |
| 130 | Elektra | 15-Jan-05 | 14:14:01 | -6.6 | -8.6 | 0.71 | 180 | 1.08 | 0.11 | -7.4 | 6.2 | -8.8 | 3.2 |
| 130 | Elektra | 12-Mar-06 | 13:38:32 | -7.2 | -8.9 | 0.77 | 300 | 1.02 | 0.14 | -6.2 | 10.5 | -9.1 | 2.8 |
| 130 | Elektra | 08-Apr-06 | 12:03:05 | -4.5 | -9.2 | 1.84 | 300 | 1.04 | 0.15 | -3.6 | 34.4 | -9.2 | 2.6 |
| 130 | Elektra | 08-Apr-06 | 12:08:40 | -4.2 | -9.4 | 1.2 | 150 | 1.05 | 0.13 | -6.5 | 9.1 | -9.5 | 2.3 |
| 130 | Elektra | 09-Apr-06 | 09:12:20 | -6.0 | -9.0 | 1.12 | 300 | 1.09 | 0.18 | -5.4 | 15.0 | -8.9 | 3.0 |
| 130 | Elektra | 11-Apr-06 | 06:01:03 | -5.4 | -8.3 | 0.76 | 360 | 1.37 | 0.13 | -6.8 | 8.1 | -8.4 | 3.9 |
| 130 | Elektra | 11-Apr-06 | 06:11:58 | -4.0 | -8.2 | 0.98 | 360 | 1.39 | 0.14 | -6.3 | 10.2 | -8.3 | 3.9 |
| 130 | Elektra | 11-Apr-06 | 06:22:16 | -5.6 | -7.6 | 0.81 | 360 | 1.42 | 0.16 | -6.0 | 11.3 | -7.9 | 4.9 |
| 130 | Elektra | 11-Apr-06 | 11:08:12 | -5.6 | -9.1 | 1.07 | 300 | 1.01 | 0.12 | -6.0 | 11.6 | -9.2 | 2.7 |
| 130 | Elektra | 12-Apr-06 | 11:56:27 | -6.8 | -8.3 | 0.77 | 140 | 1.06 | 0.13 | -6.8 | 7.9 | -8.3 | 4.0 |
| 130 | Elektra | 13-Apr-06 | 12:16:56 | -7.1 | -8.7 | 0.77 | 300 | 1.10 | 0.15 | -5.6 | 13.6 | -8.7 | 3.3 |
| 130 | Elektra | 27-Apr-06 | 03:33:18 | -4.7 | -8.1 | 0.85 | 360 | 1.36 | 0.12 | -6.5 | 9.1 | -8.1 | 4.4 |
| 130 | Elektra | 27-Apr-06 | 06:20:43 | -8.6 | -6.4 | 0.61 | 300 | 1.42 | 0.26 | -4.3 | 25.0 | -6.6 | 8.7 |
| 130 | Elektra | 28-Apr-06 | 03:39:12 | -7.5 | -8.1 | 0.62 | 360 | 1.35 | 0.12 | -7.1 | 7.0 | -8.1 | 4.3 |
| 130 | Elektra | 30-Apr-06 | 03:26:54 | -7.9 | -6.0 | 0.45 | 360 | 1.36 | 0.14 | -5.5 | 14.8 | -6.0 | 11.8 |
| 130 | Elektra | 02-May-06 | 04:48:53 | -4.0 | -6.1 | 0.82 | 360 | 1.46 | 0.17 | -4.6 | 21.9 | -6.1 | 11.0 |
| 130 | Elektra | 15-May-06 | 09:49:07 | -9.1 | -9.4 | 0.74 | 360 | 1.09 | 0.12 | -6.4 | 9.5 | -9.6 | 2.2 |
| 130 | Elektra | 16-May-06 | 08:14:44 | -8.6 | -7.6 | 0.7 | 300 | 1.00 | 0.19 | -5.3 | 15.7 | -7.9 | 4.8 |
| 130 | Elektra | 20-May-06 | 01:58:18 | -5.3 | -8.2 | 0.73 | 360 | 1.36 | 0.12 | -6.2 | 10.4 | -8.2 | 4.1 |
| 130 | Elektra | 20-May-06 | 02:08:18 | -4.5 | -8.4 | 0.85 | 360 | 1.36 | 0.12 | -6.1 | 10.8 | -8.6 | 3.5 |
| 130 | Elektra | 20-May-06 | 02:18:18 | -4.5 | -8.1 | 0.93 | 360 | 1.36 | 0.13 | -6.1 | 10.8 | -8.2 | 4.2 |
| 130 | Elektra | 23-May-06 | 03:04:13 | -6.9 | -8.7 | 0.76 | 360 | 1.43 | 0.11 | -7.3 | 6.3 | -8.7 | 3.4 |
| 130 | Elektra | 28-May-06 | 01:46:51 | -7.0 | -7.1 | 0.64 | 360 | 1.36 | 0.14 | -6.0 | 11.5 | -7.1 | 6.8 |
| 130 | Elektra | 29-May-06 | 02:30:30 | -5.4 | -6.9 | 0.69 | 360 | 1.41 | 0.17 | -5.3 | 16.2 | -6.9 | 7.7 |
| 130 | Elektra | 02-Jun-06 | 02:03:32 | -7.6 | -7.3 | 0.6 | 360 | 1.39 | 0.14 | -5.4 | 14.9 | -7.2 | 6.7 |
| 130 | Elektra | 03-Apr-07 | 14:08:23 | -6.4 | -7.1 | 0.82 | 360 | 1.08 | 0.25 | -4.1 | 28.0 | -7.2 | 6.7 |
| 283 | Emma | 15-Jul-03 | 06:55:27 | -4.7 | -9.7 | 1.46 | 286 | 1.02 | 0.17 | -6.8 | 6.3 | -9.7 | 1.7 |
| 283 | Emma | 15-Jul-03 | 07:13:31 | -7.2 | -7.1 | 0.73 | 8 | 1.03 | 0.11 | -7.0 | 6.0 | -7.1 | 5.7 |
| 283 | Emma | 15-Jul-03 | 07:17:02 | -9.7 | -7.2 | 0.49 | 14 | 1.03 | 0.11 | -7.0 | 5.8 | -7.2 | 5.4 |
| 283 | Emma | 15-Jul-03 | 07:20:20 | -4.9 | -7.5 | 0.92 | 14 | 1.03 | 0.11 | -7.1 | 5.7 | -7.5 | 4.6 |
| 283 | Emma | 16-Jul-03 | 10:02:43 | -2.1 | -5.0 | 1.36 | 57 | 1.54 | 0.36 | -3.8 | 26.2 | -5.0 | 14.8 |
| 283 | Emma | 16-Jul-03 | 10:27:27 | -2.8 | -7.0 | 1.41 | 96 | 1.77 | 0.16 | -5.0 | 15.0 | -7.0 | 6.0 |
| 283 | Emma | 30-Oct-04 | 12:16:22 | -5.4 | -9.4 | 0.99 | 70 | 1.04 | 0.14 | -5.8 | 10.2 | -9.4 | 2.0 |
| 283 | Emma | 30-Oct-04 | 14:03:46 | -6.5 | -9.3 | 1.12 | 40 | 1.07 | 0.13 | -5.2 | 13.7 | -9.4 | 2.0 |
| 283 | Emma | 30-Oct-04 | 15:23:30 | -4.8 | -8.8 | 1.36 | 40 | 1.24 | 0.15 | -5.2 | 13.4 | -8.6 | 2.8 |
| 283 | Emma | 02-Nov-04 | 15:20:11 | -5.9 | -8.9 | 0.77 | 40 | 1.27 | 0.12 | -6.8 | 6.5 | -8.9 | 2.5 |
| 283 | Emma | 05-Nov-04 | 10:30:55 | -4.0 | -9.3 | 1.29 | 50 | 1.14 | 0.12 | -6.7 | 6.7 | -9.4 | 2.0 |
| 283 | Emma | 14-Nov-04 | 06:31:31 | -6.1 | -8.6 | 0.85 | 300 | 1.86 | 0.11 | -7.7 | 4.3 | -8.6 | 2.8 |
| 283 | Emma | 15-Nov-04 | 05:42:46 | -5.9 | -8.4 | 0.88 | 300 | 1.85 | 0.12 | -7.3 | 5.2 | -8.5 | 3.0 |
| 283 | Emma | 16-Nov-04 | 04:58:46 | -6.0 | -7.8 | 0.65 | 300 | 1.93 | 0.13 | -5.9 | 9.9 | -7.8 | 4.1 |
| 283 | Emma | 16-Nov-04 | 05:56:37 | -4.6 | -8.5 | 0.88 | 300 | 1.84 | 0.11 | -6.1 | 8.8 | -8.5 | 3.0 |
| 283 | Emma | 17-Nov-04 | 05:08:30 | -9.4 | -8.7 | 0.62 | 300 | 1.89 | 0.11 | -7.9 | 3.9 | -8.7 | 2.8 |
| 283 | Emma | 18-Nov-04 | 06:19:18 | -6.6 | -8.2 | 0.84 | 300 | 1.87 | 0.13 | -6.7 | 6.7 | -8.2 | 3.4 |
| 283 | Emma | 07-Dec-04 | 03:38:55 | -5.4 | -8.6 | 0.8 | 720 | 1.84 | 0.12 | -7.1 | 5.7 | -8.7 | 2.7 |
| 283 | Emma | 07-Dec-04 | 03:55:49 | -3.5 | -8.8 | 0.98 | 720 | 1.81 | 0.13 | -6.8 | 6.5 | -8.9 | 2.5 |
| 283 | Emma | 07-Dec-04 | 04:11:39 | -3.1 | -8.4 | 1.23 | 720 | 1.80 | 0.13 | -6.4 | 7.6 | -8.5 | 3.0 |
| 283 | Emma | 08-Dec-04 | 04:17:35 | -3.1 | -7.8 | 1.05 | 300 | 1.79 | 0.12 | -6.5 | 7.5 | -7.8 | 4.1 |
| 283 | Emma | 10-Dec-04 | 05:46:52 | -4.8 | -8.9 | 0.86 | 300 | 2.10 | 0.12 | -6.7 | 6.8 | -9.0 | 2.4 |
| 283 | Emma | 14-Dec-04 | 03:55:29 | -6.7 | -7.9 | 0.64 | 720 | 1.78 | 0.12 | -6.9 | 6.1 | -8.4 | 3.1 |
| 283 | Emma | 14-Dec-04 | 04:12:42 | -7.6 | -7.7 | 0.62 | 720 | 1.79 | 0.13 | -7.0 | 5.9 | -7.9 | 4.0 |
| 283 | Emma | 14-Dec-04 | 04:32:18 | -3.4 | -8.0 | 1.11 | 720 | 1.83 | 0.14 | -6.4 | 7.8 | -8.1 | 3.6 |
| 283 | Emma | 19-Dec-04 | 01:50:48 | -5.0 | -7.0 | 0.62 | 120 | 1.26 | 0.14 | -6.2 | 8.6 | -7.0 | 5.8 |
| 283 | Emma | 19-Dec-04 | 03:29:22 | -4.5 | -8.1 | 0.85 | 300 | 1.76 | 0.13 | -6.7 | 6.7 | -8.1 | 3.6 |
| 283 | Emma | 20-Dec-04 | 01:42:25 | -3.7 | -8.1 | 0.98 | 300 | 1.96 | 0.12 | -6.1 | 9.1 | -8.1 | 3.6 |
| 283 | Emma | 20-Dec-04 | 04:32:17 | -6.5 | -8.1 | 0.66 | 300 | 1.91 | 0.12 | -6.1 | 8.9 | -8.0 | 3.7 |
| 283 | Emma | 28-Dec-04 | 02:44:43 | -6.2 | -8.5 | 0.8 | 300 | 1.73 | 0.11 | -7.3 | 5.2 | -8.5 | 2.9 |
| 283 | Emma | 28-Dec-04 | 04:52:39 | -8.0 | -8.0 | 0.65 | 300 | 2.26 | 0.11 | -6.8 | 6.4 | -8.0 | 3.8 |
| 283 | Emma | 07-Jun-06 | 05:58:31 | -10.6 | -7.6 | 0.46 | 300 | 1.34 | 0.13 | -4.7 | 16.8 | -7.6 | 4.4 |
| 283 | Emma | 11-Jun-06 | 06:10:17 | -7.9 | -7.9 | 0.61 | 300 | 1.48 | 0.11 | -4.8 | 15.9 | -7.9 | 4.0 |

**Table 3b:** Search for moonlet companions around (379) Huenna and (3749) Balam.

| ID | Name | Date | UT | α | Dmlim | r lim arcsec | Int Time s | Airmass | FWHM arcsec | Δm at 2/100xRHill | Diameter at 2/100XRHill | Δm at 1/4xRHill | Diameter at 1/4xRHill |
|---|---|---|---|---|---|---|---|---|---|---|---|---|---|
| 379 | Huenna | 05-Nov-04 | 15:20:31 | -6.1 | -8.3 | 0.85 | 180 | 1.00 | 0.13 | -4.5 | 11.8 | -8.3 | 2.0 |
| 379 | Huenna | 07-Dec-04 | 09:11:39 | -6.5 | -6.5 | 0.53 | 180 | 1.57 | 0.1 | -4.6 | 11.2 | -6.5 | 4.7 |
| 379 | Huenna | 08-Dec-04 | 07:08:41 | -4.0 | -7.0 | 0.78 | 300 | 1.38 | 0.09 | -4.7 | 10.8 | -7.0 | 3.7 |
| 379 | Huenna | 09-Dec-04 | 06:35:44 | -4.9 | -7.7 | 0.78 | 300 | 1.41 | 0.08 | -5.7 | 6.6 | -7.6 | 2.8 |
| 379 | Huenna | 09-Dec-04 | 06:48:16 | -6.0 | -8.1 | 0.70 | 300 | 1.40 | 0.08 | -6.2 | 5.3 | -8.1 | 2.2 |
| 379 | Huenna | 10-Dec-04 | 06:51:34 | -8.4 | -8.3 | 0.53 | 300 | 1.39 | 0.08 | -6.7 | 4.2 | -8.3 | 2.1 |
| 379 | Huenna | 14-Dec-04 | 05:28:48 | -6.4 | -7.7 | 0.58 | 300 | 1.51 | 0.08 | -5.9 | 6.2 | -7.7 | 2.7 |
| 379 | Huenna | 14-Dec-04 | 07:09:01 | -5.7 | -7.6 | 0.64 | 300 | 1.39 | 0.08 | -5.9 | 6.1 | -7.6 | 2.8 |
| 379 | Huenna | 15-Dec-04 | 05:20:30 | -7.4 | -7.8 | 0.53 | 300 | 1.53 | 0.08 | -5.8 | 6.3 | -7.7 | 2.6 |
| 379 | Huenna | 28-Dec-04 | 05:37:03 | -6.1 | -8.0 | 0.66 | 300 | 1.39 | 0.1 | -5.6 | 7.0 | -8.0 | 2.3 |
| 379 | Huenna | 28-Dec-04 | 07:41:22 | -4.2 | -7.7 | 0.74 | 300 | 1.61 | 0.1 | -5.5 | 7.2 | -7.7 | 2.7 |
| 379 | Huenna | 29-Dec-04 | 05:13:41 | -6.3 | -8.3 | 0.57 | 300 | 1.41 | 0.08 | -6.2 | 5.4 | -8.3 | 2.1 |
| 379 | Huenna | 18-Jan-05 | 03:58:39 | -5.3 | -8.2 | 0.52 | 300 | 1.41 | 0.08 | -6.3 | 5.2 | -8.1 | 2.2 |
| 379 | Huenna | 18-Jan-05 | 06:17:38 | -5.5 | -7.8 | 0.56 | 300 | 1.71 | 0.09 | -5.8 | 6.4 | -7.8 | 2.6 |
| 379 | Huenna | 21-Jan-05 | 02:25:32 | -4.5 | -7.2 | 0.72 | 300 | 1.56 | 0.09 | -5.1 | 9.0 | -7.2 | 3.4 |
| 379 | Huenna | 25-Jan-05 | 04:51:45 | -3.6 | -6.7 | 0.78 | 300 | 1.51 | 0.11 | -4.8 | 10.0 | -6.6 | 4.4 |
| 379 | Huenna | 25-Jan-05 | 06:43:58 | -2.2 | -5.6 | 1.04 | 300 | 2.26 | 0.18 | -3.8 | 15.8 | -5.6 | 7.1 |
| 379 | Huenna | 26-Jan-05 | 02:47:49 | -4.2 | -8.0 | 0.72 | 300 | 1.45 | 0.08 | -6.0 | 5.8 | -8.0 | 2.3 |
| 379 | Huenna | 26-Jan-05 | 05:10:53 | -5.3 | -7.6 | 0.58 | 300 | 1.58 | 0.09 | -5.8 | 6.5 | -7.6 | 2.8 |
| 379 | Huenna | 27-Jan-05 | 03:10:56 | -6.1 | -7.3 | 0.56 | 300 | 1.42 | 0.09 | -5.7 | 6.8 | -7.3 | 3.3 |
| 379 | Huenna | 27-Jan-05 | 06:08:04 | -3.9 | -5.9 | 0.86 | 300 | 1.97 | 0.12 | -3.2 | 21.4 | -5.9 | 6.0 |
| 379 | Huenna | 28-Jan-05 | 03:04:48 | -4.6 | -7.9 | 0.81 | 300 | 1.42 | 0.09 | -5.3 | 8.1 | -7.9 | 2.4 |
| 379 | Huenna | 28-Jan-05 | 03:14:05 | -4.4 | -8.0 | 0.84 | 300 | 1.42 | 0.1 | -5.3 | 8.2 | -8.0 | 2.4 |
| 379 | Huenna | 28-Jan-05 | 03:22:34 | -4.0 | -6.4 | 0.68 | 300 | 1.41 | 0.17 | -4.1 | 14.0 | -6.4 | 4.9 |
| 379 | Huenna | 02-Feb-05 | 03:09:22 | -6.9 | -8.1 | 0.52 | 300 | 1.42 | 0.08 | -6.0 | 5.9 | -8.1 | 2.3 |
| 379 | Huenna | 02-Feb-05 | 05:09:40 | -3.8 | -7.7 | 0.82 | 300 | 1.74 | 0.1 | -5.3 | 8.2 | -7.7 | 2.7 |
| 379 | Huenna | 04-Feb-05 | 02:41:11 | -4.9 | -7.9 | 0.45 | 300 | 1.42 | 0.08 | -6.5 | 4.7 | -7.8 | 2.5 |
| 379 | Huenna | 04-Feb-05 | 04:06:03 | -7.4 | -7.8 | 0.49 | 300 | 1.51 | 0.09 | -5.6 | 7.1 | -7.8 | 2.6 |
| 379 | Huenna | 04-Feb-05 | 04:14:50 | -5.0 | -7.3 | 0.98 | 300 | 1.54 | 0.1 | -3.4 | 19.2 | -7.3 | 3.2 |
| 379 | Huenna | 04-Feb-05 | 04:23:59 | -5.0 | -6.6 | 0.84 | 300 | 1.57 | 0.12 | -4.0 | 14.5 | -6.6 | 4.3 |
| 379 | Huenna | 07-Feb-05 | 03:44:25 | -6.1 | -7.1 | 0.60 | 300 | 1.50 | 0.1 | -4.9 | 9.6 | -7.1 | 3.5 |
| 379 | Huenna | 08-Feb-05 | 02:30:20 | -3.5 | -6.9 | 0.81 | 600 | 1.42 | 0.1 | -4.8 | 10.0 | -7.0 | 3.7 |
| 379 | Huenna | 08-Feb-05 | 02:45:38 | -4.0 | -7.0 | 0.72 | 300 | 1.42 | 0.09 | -5.1 | 9.0 | -6.9 | 3.8 |
| 379 | Huenna | 09-Feb-05 | 03:16:53 | -4.7 | -7.4 | 0.48 | 300 | 1.46 | 0.09 | -5.9 | 6.0 | -7.3 | 3.1 |
| 379 | Huenna | 16-Feb-05 | 01:21:14 | -5.5 | -7.7 | 0.49 | 300 | 1.45 | 0.08 | -6.1 | 5.6 | -7.7 | 2.6 |
| 3749 | Balam | 15-Jul-03 | 05:28:54 | -15.9 | -5.8 | 0.30 | 600 | 1.07 | 0.12 | -1.8 | 2.7 | -3.8 | 1.1 |
| 3749 | Balam | 16-Jul-03 | 04:22:13 | -19.3 | -6.0 | 0.30 | 160 | 1.01 | 0.11 | -1.4 | 3.3 | -3.6 | 1.2 |
| 3749 | Balam | 14-Nov-04 | 06:03:30 | -9.5 | -8.1 | 0.60 | 1200 | 1.76 | 0.12 | -1.4 | 3.3 | -6.2 | 0.4 |
| 3749 | Balam | 15-Nov-04 | 03:38:09 | -12.0 | -8.1 | 0.46 | 1200 | 1.82 | 0.1 | -1.0 | 4.0 | -6.0 | 0.4 |
| 3749 | Balam | 15-Nov-04 | 04:05:35 | -8.3 | -8.1 | 0.76 | 1200 | 1.73 | 0.13 | -1.3 | 3.4 | -5.6 | 0.5 |
| 3749 | Balam | 16-Nov-04 | 05:30:58 | -7.0 | -7.1 | 0.62 | 1200 | 1.70 | 0.15 | -1.5 | 3.1 | -4.9 | 0.7 |
| 3749 | Balam | 17-Nov-04 | 04:39:31 | -9.0 | -7.9 | 0.62 | 1200 | 1.67 | 0.12 | -1.8 | 2.7 | -6.0 | 0.4 |
| 3749 | Balam | 22-Nov-04 | 03:09:25 | -5.2 | -6.5 | 0.98 | 1200 | 1.76 | 0.22 | -2.0 | 2.4 | -3.9 | 1.1 |
| 3749 | Balam | 02-Dec-04 | 03:50:18 | -5.6 | -6.4 | 0.79 | 1200 | 1.62 | 0.19 | -2.0 | 2.5 | -3.8 | 1.1 |
| 3749 | Balam | 03-Dec-04 | 04:02:53 | -8.0 | -6.6 | 0.68 | 1200 | 1.63 | 0.23 | -3.6 | 1.2 | -3.1 | 1.5 |
| 3749 | Balam | 07-Dec-04 | 03:02:08 | -2.9 | -5.6 | 0.84 | 1200 | 1.59 | 0.17 | -2.3 | 2.2 | -3.3 | 1.3 |
| 3749 | Balam | 09-Dec-04 | 03:32:49 | -3.7 | -6.0 | 0.84 | 1200 | 1.61 | 0.14 | -1.9 | 2.6 | -3.5 | 1.2 |
| 3749 | Balam | 10-Dec-04 | 02:44:19 | -5.8 | -7.3 | 0.79 | 1200 | 1.58 | 0.11 | -1.0 | 4.0 | -4.1 | 1.0 |
| 3749 | Balam | 14-Dec-04 | 02:48:59 | -4.6 | -5.5 | 0.62 | 720 | 1.57 | 0.32 | -3.0 | 1.6 | -2.7 | 1.8 |
| 3749 | Balam | 14-Dec-04 | 03:25:21 | -6.1 | -6.3 | 0.81 | 1200 | 1.61 | 0.21 | -2.1 | 2.4 | -3.1 | 1.5 |
| 3749 | Balam | 20-Dec-04 | 01:12:00 | -4.6 | -6.3 | 0.70 | 1200 | 1.60 | 0.13 | -2.1 | 2.4 | -3.6 | 1.2 |
| 3749 | Balam | 20-Dec-04 | 03:57:54 | -4.2 | -6.1 | 0.73 | 1200 | 1.80 | 0.15 | -1.9 | 2.6 | -3.8 | 1.1 |

**Table 4a:**
Size, shape and orientation of Elektra's primary and comparison with Durech *et al.* (2006) model with a pole solution (λ= 68°, β= -88°) in EC2000 and $P_{spin}$ = 5.224 h. The average diameter of (130) Elektra ($D_{AO}$ = 215 ± 15 km) is 16% larger than STM radiometric measurement (Tedesco *et al.* 2002).

| ID  | Name    | Date       | UT       | 2a (mas) | 2b (mas) | 2a (km) | 2b (km) | Observed Orientation (deg) | a/b  | DAO (km) |
|-----|---------|------------|----------|----------|----------|---------|---------|----------------------------|------|----------|
| 130 | Elektra | 7-Dec-03   | 07:16:10 | 178±2    | 138±2    | 224±5   | 174±2   | 10                         | 1.29 | 199      |
| 130 | Elektra | 5-Jan-04   | 02:59:13 | 113±6    | 103±6    | 162±8   | 147±9   | 63                         | 1.10 | 155      |
| 130 | Elektra | 5-Jan-04   | 04:25:39 | 138±4    | 97±6     | 198±6   | 140±9   | -22                        | 1.42 | 169      |
| 130 | Elektra | 6-Jan-04   | 03:06:56 | 132±5    | 112±6    | 190±7   | 162±8   | 9                          | 1.17 | 177      |
| 130 | Elektra | 15-Jan-05  | 12:25:31 | 136±2    | 83±4     | 246±3   | 150±8   | 23                         | 1.63 | 198      |
| 130 | Elektra | 15-Jan-05  | 14:14:01 | 124±2    | 94±4     | 225±4   | 170±7   | 21                         | 1.32 | 197      |
| 130 | Elektra | 11-Apr-06  | 06:01:03 | 139±4    | 98±6     | 287±9   | 202±13  | 26                         | 1.42 | 245      |
| 130 | Elektra | 11-Apr-06  | 06:11:58 | 156±4    | 109±6    | 322±8   | 226±12  | 23                         | 1.43 | 274      |
| 130 | Elektra | 27-Apr-06  | 03:33:18 | 126±5    | 85±7     | 265±11  | 179±14  | 40                         | 1.48 | 222      |
| 130 | Elektra | 28-Apr-06  | 03:39:12 | 130±5    | 102±6    | 274±10  | 215±13  | 10                         | 1.27 | 245      |
| 130 | Elektra | 20-May-06  | 01:58:18 | 135±5    | 96±6     | 298±10  | 215±14  | 10                         | 1.41 | 256      |
| 130 | Elektra | 23-May-06  | 03:04:13 | 123±5    | 96±6     | 277±12  | 218±14  | 9                          | 1.27 | 248      |

**Table 4b:**
Size, shape and orientation of Emma's primary. The AO images were fitted by an ellipse function defined by its major axes (2a, 2b) and its orientation (from the celestial east, and counter-clockwise). The a/b ratio and the average diameter ($D_{avg}$) are also labeled. The average diameter of (283) Emma ($D_{AO} = 160 \pm 10$ km) is 8% larger than STM/IRAS radiometric measurement (Tedesco *et al.* 2002).

| ID | Name | Date | UT | 2a (mas) | 2b (mas) | 2a (km) | 2b (km) | Observed Orientation (deg) | a/b | DAO (km) |
|---|---|---|---|---|---|---|---|---|---|---|
| 283 | Emma | 30-Oct-04 | 12:16:22 | 158±10 | 126±12 | 227±15 | 181±17 | -22 | 1.25 | 204 |
| 283 | Emma | 30-Oct-04 | 14:03:46 | 161±10 | 144±11 | 232±15 | 208±16 | 38 | 1.12 | 220 |
| 283 | Emma | 30-Oct-04 | 15:23:30 | 177±10 | 134±11 | 254±14 | 193±16 | -4 | 1.32 | 223 |
| 283 | Emma | 02-Nov-04 | 15:20:11 | 142±11 | 127±12 | 202±16 | 180±16 | -45 | 1.12 | 191 |
| 283 | Emma | 05-Nov-04 | 10:30:55 | 115±12 | NA | 162±17 | NA | 0 | NA | NA |
| 283 | Emma | 14-Nov-04 | 06:31:31 | 120±5 | 93±6 | 164±7 | 129±9 | -42 | 1.28 | 146 |
| 283 | Emma | 15-Nov-04 | 05:42:46 | 112±6 | 90±6 | 154±8 | 123±9 | 61 | 1.25 | 139 |
| 283 | Emma | 16-Nov-04 | 04:58:46 | 126±5 | 107±6 | 173±7 | 146±8 | 22 | 1.18 | 160 |
| 283 | Emma | 16-Nov-04 | 05:56:37 | 107±6 | 94±6 | 147±8 | 129±9 | 86 | 1.14 | 138 |
| 283 | Emma | 17-Nov-04 | 05:08:30 | 112±6 | 90±6 | 153±8 | 123±9 | 24 | 1.25 | 138 |
| 283 | Emma | 18-Nov-04 | 06:19:18 | 115±5 | 107±6 | 157±7 | 147±8 | 87 | 1.07 | 152 |
| 283 | Emma | 07-Dec-04 | 03:38:55 | 128±5 | 94±6 | 174±7 | 128±9 | -17 | 1.36 | 151 |
| 283 | Emma | 07-Dec-04 | 03:55:49 | 122±5 | 109±6 | 166±7 | 149±8 | 0 | 1.11 | 157 |
| 283 | Emma | 07-Dec-04 | 04:11:39 | 126±5 | 116±5 | 172±7 | 157±7 | 21 | 1.09 | 165 |
| 283 | Emma | 08-Dec-04 | 04:17:35 | 121±5 | 103±6 | 166±7 | 140±8 | 19 | 1.18 | 153 |
| 283 | Emma | 10-Dec-04 | 05:46:52 | 130±5 | 107±6 | 178±7 | 146±8 | 82 | 1.22 | 162 |
| 283 | Emma | 14-Dec-04 | 03:55:29 | 123±5 | 92±6 | 170±7 | 127±9 | -40 | 1.34 | 149 |
| 283 | Emma | 14-Dec-04 | 04:12:42 | 130±5 | 113±6 | 180±7 | 156±8 | -34 | 1.15 | 168 |
| 283 | Emma | 14-Dec-04 | 04:32:18 | 142±4 | 131±5 | 196±6 | 180±7 | 8 | 1.09 | 188 |
| 283 | Emma | 19-Dec-04 | 03:29:22 | 128±5 | 98±6 | 179±7 | 137±9 | -90 | 1.31 | 158 |
| 283 | Emma | 20-Dec-04 | 01:42:25 | 130±5 | 100±6 | 182±7 | 141±8 | -27 | 1.29 | 161 |
| 283 | Emma | 20-Dec-04 | 04:32:17 | 122±5 | 97±6 | 171±7 | 136±9 | -42 | 1.26 | 153 |
| 283 | Emma | 28-Dec-04 | 02:44:43 | 127±5 | 90±6 | 183±7 | 130±9 | -29 | 1.41 | 156 |
| 283 | Emma | 28-Dec-04 | 04:52:39 | 107±6 | 98±6 | 155±8 | 142±9 | -260 | 1.09 | 148 |

**Table 5a:**

Characteristics of the moonlet of (130) Elektra (named S/2003(130)1) measured on the AO images collected with VLT-UT4, Gemini and Keck in 2004-2006. The X and Y relative positions with respect to the primary of the system are measured by fitting their centroid profile with a Moffat-Gauss function. The diameter of satellite is estimated by calculating the integrated flux ratio of the primary and the secondary and also by measuring directly the diameter size of the primary on the resolved AO images (see Table 4a).

| ID | Primary Name | Date | UT | Telescope | X | Y | separation | Δm (peak-to-peak) | Δm (integrated) | Satellite Size |
|---|---|---|---|---|---|---|---|---|---|---|
| | | | | | arcsec | arcsec | arcsec | | | km |
| 130 | Elektra | 07-Dec-03 | 7:16:10 | Keck | -0.570 | -0.568 | 0.805 | -5.43 | -7.90 | 5.3 |
| 130 | Elektra | 05-Jan-04 | 02:59:13 | VLT | 0.903 | 0.293 | 0.949 | -6.06 | -6.82 | 8.1 |
| 130 | Elektra | 05-Jan-04 | 04:25:39 | VLT | 0.866 | 0.315 | 0.921 | -6.01 | -7.27 | 7.3 |
| 130 | Elektra | 06-Jan-04 | 03:06:56 | VLT | 0.013 | 0.395 | 0.395 | -4.91 | -7.16 | 7.1 |
| 130 | Elektra | 02-Mar-04 | 00:30:54 | VLT | 0.502 | -0.104 | 0.513 | -3.59 | -7.73 | 12.7 |
| 130 | Elektra | 15-Jan-05 | 12:25:31 | Keck | -0.645 | 0.249 | 0.691 | -5.44 | -8.51 | 4.3 |
| 130 | Elektra | 15-Jan-05 | 14:14:01 | Keck | -0.626 | 0.248 | 0.673 | -5.66 | -8.73 | 3.8 |
| 130 | Elektra | 08-Apr-06 | 12:08:40 | Gemini | -0.654 | 0.216 | 0.689 | -5.56 | -7.65 | 7.8 |
| 130 | Elektra | 11-Apr-06 | 06:01:03 | VLT | 0.516 | -0.199 | 0.553 | -5.20 | -8.17 | 6.2 |
| 130 | Elektra | 11-Apr-06 | 06:11:58 | VLT | 0.529 | -0.212 | 0.569 | -4.58 | -8.66 | 5.4 |
| 130 | Elektra | 23-May-06 | 03:04:13 | VLT | 0.357 | 0.013 | 0.357 | -5.46 | -7.73 | 7.1 |

**Table 5b:**

Characteristic of the moonlet orbiting around (283) Emma (named S/200X(283) 1) measured on the AO images collected with VLT-UT4 and Gemini in 2003-2004. The X and Y relative positions with respect to the primary of the system is measured fitting their centroid profile with a Moffat-Gauss function. The diameter of satellite is estimated by calculating the integrated flux ratio of the primary and the secondary and also by measuring directly the same of the primary on the resolved AO images (see Table 4b).

| ID | Primary Name | Date | UT | Telescope | X | Y | separation | Δm (peak-to-peak) | Δm (integrated) | Satellite Size |
|---|---|---|---|---|---|---|---|---|---|---|
| | | | | | arcsec | arcsec | arcsec | | | km |
| 283 | Emma | 7/15/03 | 07:13:31 | VLT | 0.095 | -0.367 | 0.379 | -3.58 | -4.12 | 21.53 |
| 283 | Emma | 7/15/03 | 07:17:02 | VLT | 0.105 | -0.375 | 0.389 | -3.71 | -4.27 | 19.07 |
| 283 | Emma | 7/15/03 | 07:20:20 | VLT | 0.098 | -0.356 | 0.369 | -3.59 | -4.09 | 21.27 |
| 283 | Emma | 7/16/03 | 10:27:27 | VLT | 0.136 | 0.373 | 0.397 | -2.54 | -6.25 | 11.37 |
| 283 | Emma | 10/30/04 | 12:16:22 | Gemini | 0.441 | -0.024 | 0.442 | -3.58 | -5.09 | 18.46 |
| 283 | Emma | 11/2/04 | 15:20:11 | Gemini | 0.417 | 0.139 | 0.439 | -3.55 | -4.46 | 22.38 |
| 283 | Emma | 11/14/04 | 06:31:31 | VLT | -0.251 | -0.258 | 0.360 | -3.54 | -4.23 | 22.48 |
| 283 | Emma | 11/15/04 | 05:42:46 | VLT | -0.118 | 0.343 | 0.363 | -3.52 | -4.78 | 17.78 |
| 283 | Emma | 11/16/04 | 04:58:46 | VLT | 0.451 | 0.130 | 0.470 | -3.46 | -5.25 | 16.03 |
| 283 | Emma | 11/16/04 | 05:56:37 | VLT | 0.455 | 0.119 | 0.471 | -3.65 | -4.33 | 22.02 |
| 283 | Emma | 11/17/04 | 05:08:30 | VLT | 0.095 | -0.356 | 0.369 | -3.55 | -4.25 | 22.60 |
| 283 | Emma | 11/18/04 | 06:19:18 | VLT | -0.330 | 0.158 | 0.366 | -3.37 | -5.69 | 12.80 |
| 283 | Emma | 12/7/04 | 03:38:55 | VLT | 0.348 | -0.226 | 0.415 | -3.45 | -4.80 | 18.25 |
| 283 | Emma | 12/7/04 | 03:55:49 | VLT | 0.345 | -0.237 | 0.419 | -3.26 | -5.48 | 13.67 |
| 283 | Emma | 12/7/04 | 04:11:39 | VLT | 0.340 | -0.237 | 0.414 | -3.24 | -5.46 | 14.93 |
| 283 | Emma | 12/8/04 | 04:17:35 | VLT | -0.343 | -0.160 | 0.378 | -3.19 | -5.16 | 15.74 |
| 283 | Emma | 12/10/04 | 05:46:52 | VLT | 0.450 | -0.062 | 0.455 | -3.55 | -4.57 | 20.64 |
| 283 | Emma | 12/14/04 | 03:55:29 | VLT | 0.199 | -0.329 | 0.385 | -3.54 | -4.71 | 18.80 |
| 283 | Emma | 12/14/04 | 04:12:42 | VLT | 0.198 | -0.330 | 0.385 | -3.23 | -5.51 | 13.76 |
| 283 | Emma | 12/14/04 | 04:32:18 | VLT | 0.184 | -0.328 | 0.376 | -2.98 | -5.36 | 16.63 |
| 283 | Emma | 12/19/04 | 03:29:22 | VLT | -0.079 | 0.344 | 0.353 | -3.41 | -5.37 | 14.92 |
| 283 | Emma | 12/20/04 | 01:42:25 | VLT | 0.438 | 0.154 | 0.464 | -3.56 | -4.85 | 18.01 |
| 283 | Emma | 12/20/04 | 04:32:17 | VLT | 0.451 | 0.087 | 0.459 | -3.55 | -4.74 | 18.86 |
| 283 | Emma | 12/28/04 | 02:44:43 | VLT | -0.100 | -0.341 | 0.356 | -3.52 | -4.25 | 23.18 |
| 283 | Emma | 12/28/04 | 04:52:39 | VLT | -0.159 | -0.307 | 0.346 | -3.44 | -4.69 | 18.96 |

**Table 5c**: Characteristics of the moonlet of 379 Huenna (named S/2003 (379) 1) measured on the AO images collected with VLT/NACO in 2004-2005. The X and Y relative positions with respect to the primary of the system is measured by fitting their centroid profile with a Moffat-Gauss function. Since Huenna's primary is not resolved we estimated the moonlet diameter size using the radiometric IRAS diameter ($D_{STM}$=92.3 km).

| ID | Primary Name | Date | UT | Telescope | X | Y | separation | Δm (peak-to-peak) | Δm (integrated) | Satellite Size |
|---|---|---|---|---|---|---|---|---|---|---|
| | | | | | arcsec | arcsec | arcsec | | | km |
| 379 | Huenna | 8-Dec-04 | 07:08:41 | VLT | 1.781 | 0.125 | 1.786 | -6.08 | -8.66 | 5.62 |
| 379 | Huenna | 9-Dec-04 | 06:35:44 | VLT | 1.748 | 0.137 | 1.753 | -6.31 | -7.31 | 5.05 |
| 379 | Huenna | 9-Dec-04 | 06:48:16 | VLT | 1.739 | 0.144 | 1.745 | -6.11 | -6.86 | 5.52 |
| 379 | Huenna | 10-Dec-04 | 06:51:34 | VLT | 1.702 | 0.180 | 1.711 | -6.04 | -6.56 | 5.70 |
| 379 | Huenna | 14-Dec-04 | 05:28:48 | VLT | 1.445 | 0.262 | 1.469 | -6.25 | -7.27 | 5.18 |
| 379 | Huenna | 14-Dec-04 | 07:09:01 | VLT | 1.436 | 0.263 | 1.459 | -6.11 | -7.18 | 5.53 |
| 379 | Huenna | 15-Dec-04 | 05:20:30 | VLT | 1.373 | 0.292 | 1.404 | -6.21 | -7.67 | 5.28 |
| 379 | Huenna | 28-Dec-04 | 05:37:03 | VLT | -0.014 | 0.370 | 0.370 | -4.98 | -8.15 | 9.30 |
| 379 | Huenna | 29-Dec-04 | 05:13:41 | VLT | -0.145 | 0.383 | 0.409 | -5.29 | -6.99 | 8.08 |
| 379 | Huenna | 18-Jan-05 | 03:58:39 | VLT | -1.923 | 0.059 | 1.923 | -6.32 | -7.13 | 5.02 |
| 379 | Huenna | 18-Jan-05 | 06:17:38 | VLT | -1.922 | 0.056 | 1.923 | -6.28 | -7.69 | 5.12 |
| 379 | Huenna | 21-Jan-05 | 02:25:32 | VLT | -1.987 | 0.004 | 1.987 | -6.49 | -8.80 | 4.64 |
| 379 | Huenna | 25-Jan-05 | 04:51:45 | VLT | -1.928 | -0.111 | 1.931 | -5.89 | -8.88 | 6.14 |
| 379 | Huenna | 26-Jan-05 | 02:47:49 | VLT | -1.871 | -0.125 | 1.875 | -6.01 | -7.06 | 5.79 |
| 379 | Huenna | 26-Jan-05 | 05:10:53 | VLT | -1.875 | -0.128 | 1.879 | -6.43 | -8.67 | 4.78 |
| 379 | Huenna | 27-Jan-05 | 03:10:56 | VLT | -1.794 | -0.147 | 1.800 | -6.19 | -7.53 | 5.34 |
| 379 | Huenna | 28-Jan-05 | 03:04:48 | VLT | -1.734 | -0.172 | 1.742 | -6.02 | -7.37 | 5.77 |
| 379 | Huenna | 28-Jan-05 | 03:14:05 | VLT | -1.737 | -0.163 | 1.744 | -5.92 | -7.25 | 6.03 |
| 379 | Huenna | 28-Jan-05 | 03:22:34 | VLT | -1.737 | -0.159 | 1.744 | -5.60 | -8.41 | 7.00 |
| 379 | Huenna | 2-Feb-05 | 03:09:22 | VLT | -1.153 | -0.219 | 1.173 | -6.38 | -6.90 | 4.88 |
| 379 | Huenna | 2-Feb-05 | 05:09:40 | VLT | -1.138 | -0.226 | 1.160 | -6.11 | -7.39 | 5.55 |
| 379 | Huenna | 4-Feb-05 | 02:41:11 | VLT | -0.823 | -0.223 | 0.852 | -6.23 | -7.11 | 5.25 |
| 379 | Huenna | 4-Feb-05 | 04:06:03 | VLT | -0.823 | -0.226 | 0.854 | -6.32 | -7.35 | 5.03 |
| 379 | Huenna | 4-Feb-05 | 04:23:59 | VLT | -0.834 | -0.225 | 0.864 | -5.12 | -8.10 | 8.73 |
| 379 | Huenna | 16-Feb-05 | 01:21:14 | VLT | 1.223 | -0.006 | 1.223 | -6.23 | -7.09 | 5.23 |

**Table 5d:** Characteristics of the moonlet of 3749 Balam (named S/2002 (3749) 1) measured on the AO images collected with VLT/NACO in 2004-2005. The X and Y relative positions with respect to the primary of the system are measured by fitting their centroid profile with a Moffat-Gauss function. In Jul. 2003 and Nov. 2004, the satellite is very close to the primary limiting its detection and preventing the measurement of its flux. Since Balam's primary is not resolved, we derived the moonlet diameter using an estimated diameter for the primary ($D_p \sim 12$ km).

| ID | Primary Name | Date | UT | Telescope | X | Y | separation | Δm (peak-to-peak) | Δm (integrated) | Satellite Size |
|---|---|---|---|---|---|---|---|---|---|---|
| | | | | | arcsec | arcsec | arcsec | | | km |
| 3749 | Balam | 14-Nov-04 | 6:03:30 | VLT | 0.068 | 0.013 | 0.069 | N/A | N/A | N/A |
| 3749 | Balam | 22-Nov-04 | 3:09:25 | VLT | -0.108 | 0.054 | 0.121 | N/A | N/A | N/A |
| 3749 | Balam | 15-Jul-03 | 5:30:15 | VLT | -0.081 | -0.012 | 0.082 | N/A | N/A | N/A |
| 3749 | Balam | 16-Jul-03 | 4:22:13 | VLT | -0.081 | -0.020 | 0.083 | N/A | N/A | N/A |
| 3749 | Balam | 03-Dec-04 | 04:02:53 | VLT | -0.315 | 0.083 | 0.326 | -1.35 | -3.42 | 3.4 |
| 3749 | Balam | 07-Dec-04 | 03:02:08 | VLT | -0.321 | 0.106 | 0.338 | -1.50 | -5.68 | 3.2 |
| 3749 | Balam | 09-Dec-04 | 03:32:49 | VLT | -0.348 | 0.054 | 0.352 | -1.80 | -6.35 | 2.9 |
| 3749 | Balam | 10-Dec-04 | 02:44:19 | VLT | -0.371 | 0.078 | 0.379 | -2.38 | -4.03 | 2.3 |
| 3749 | Balam | 14-Dec-04 | 03:25:21 | VLT | -0.372 | 0.055 | 0.376 | -1.90 | -4.23 | 2.8 |
| 3749 | Balam | 20-Dec-04 | 01:12:00 | VLT | -0.372 | 0.053 | 0.375 | -2.20 | -4.91 | 2.5 |
| 3749 | Balam | 20-Dec-04 | 03:57:54 | VLT | -0.345 | 0.054 | 0.349 | -2.02 | -4.99 | 2.6 |

**Table 6:** Best-fitted orbital elements of the asteroidal companions of (130) Elektra, (283) Emma, (379) Huenna, and (3749) Balam. The orbits of the satellite and its relative location with respect to the primary is displayed in Fig 5. The orbital elements of 3749 are not well constrained. We selected an orbital solution for which the predicted satellite position is too close to the primary to be detected on 6 runs (see Section 3.2).

|  | S/2003(130)1 | S/2003(283)1 | S/2002(379)1 | S/2001(3749)1 |
|---|---|---|---|---|
| Period (days) | 5.2575 ± 0.0053 | 3.35337 ± 0.00093 | 87.60 ± 0.026 | 61 ± 10 |
| Semi-major axis (km) | 1318 ± 25 | 581.0 ± 3.6 | 3335.8 ± 54.9 | 289 ± 13 |
| Eccentricity | 0.13 ± 0.03 | 0.12 ± 0.01 | 0.222 ± 0.006 | ~0.9 |
| Inclination in J2000 (degree) | 25 ± 2 | 94.2 ± 0.4 | 152.7 ± 0.3 | unk. |
| Pericenter argument (degrees) | 311 ± 5 | 40 ± 4 | 284 ± 5 | unk |
| Time of pericenter (Julian days) | 2453834.5 ± 0.6 | 2453320.9009 ± 0.1360 | 2453326.3655 ± 0.0432 | unk. |
| Ascending Node (degrees) | 1.6 ± 2.0 | 345.4 ± 0.4 | 204.3 ± 0.3 | unk. |

**Table 7:** Physical properties of the binary asteroidal systems

|  | S/2003(130)1 | S/2003(283)1 | S/2002(379)1 | S/2001(3749)1 |
|---|---|---|---|---|
| Mass System (kg) | $6.6 \pm 0.4 \times 10^{18}$ | $1.38 \pm 0.03 \times 10^{18}$ | $3.83 \pm 0.19 \times 10^{17}$ | $5.1 \pm 0.2 \times 10^{14}$ |
| $R_{Hill}$ (km) | 58 000 | 28 000 | 20 000 | 1500 |
| a in $R_{Hill}$ | $1/40 \times R_{Hill}$ | $5/100 \times R_{Hill}$ | $1/6 \times R_{Hill}$ | $\sim 1/5 \times R_{Hill}$ |
| a in $R_{avg}^{1}$ | $14 \times R_p$ | $8 \times R_p$ | $70 \times R_p$ | $\sim 40 \times R_p$ |
| $R_{satellite}/R_{primary}$ | 0.04 | 0.06 | 0.06 | 0.43 |
| Density (g/cm$^3$) of Primary with $D_{STM/NEATM}$ | $2.1/1.7 \pm 0.3$ | $0.8/0.9 \pm 0.1$ | $0.9/0.8 \pm 0.1$ | $\sim 2.6$ |
| Density (g/cm$^3$) of Primary with $D_{AO}$ | $1.3 \pm 0.3$ | $0.7 \pm 0.2$ | Not resolved | Not resolved |
| Spin Pole Solution in ECJ2000 and degrees | 277° ± 2°<br>+85° ± 2° | 253°± 0.2°<br>+13.2°± 0.3 | 171.3°± 0.2°<br>-78.9°± 0.3 | 149.9°± 0.2°<br>+74.3°± 0.3 |

**Figure 1a:** Search for moonlets around (130) Elecktra. On the left-top figure an observation of 130 Elektra taken on Jan. 06 2004 is displayed. The right top figure corresponds to the same observations after subtracting its azimuthal average. The detection of the moonlet companion is easier. The plot below is the azimuthally averaged 2-σ detection function for this observation. It is approximated using two linear functions which depends of three parameters: α, the coefficient of the slope of the linear regimes, $r_{lim}$ the separation between 2 regimes, $\Delta m_{lim}$, the difference in magnitude in the stable regime. The minimum size of a moonlet to be detected can be derived from this profile. Table 3a contains the characteristics of the 2-σ detection curve profile for all observations of (130) Elektra.

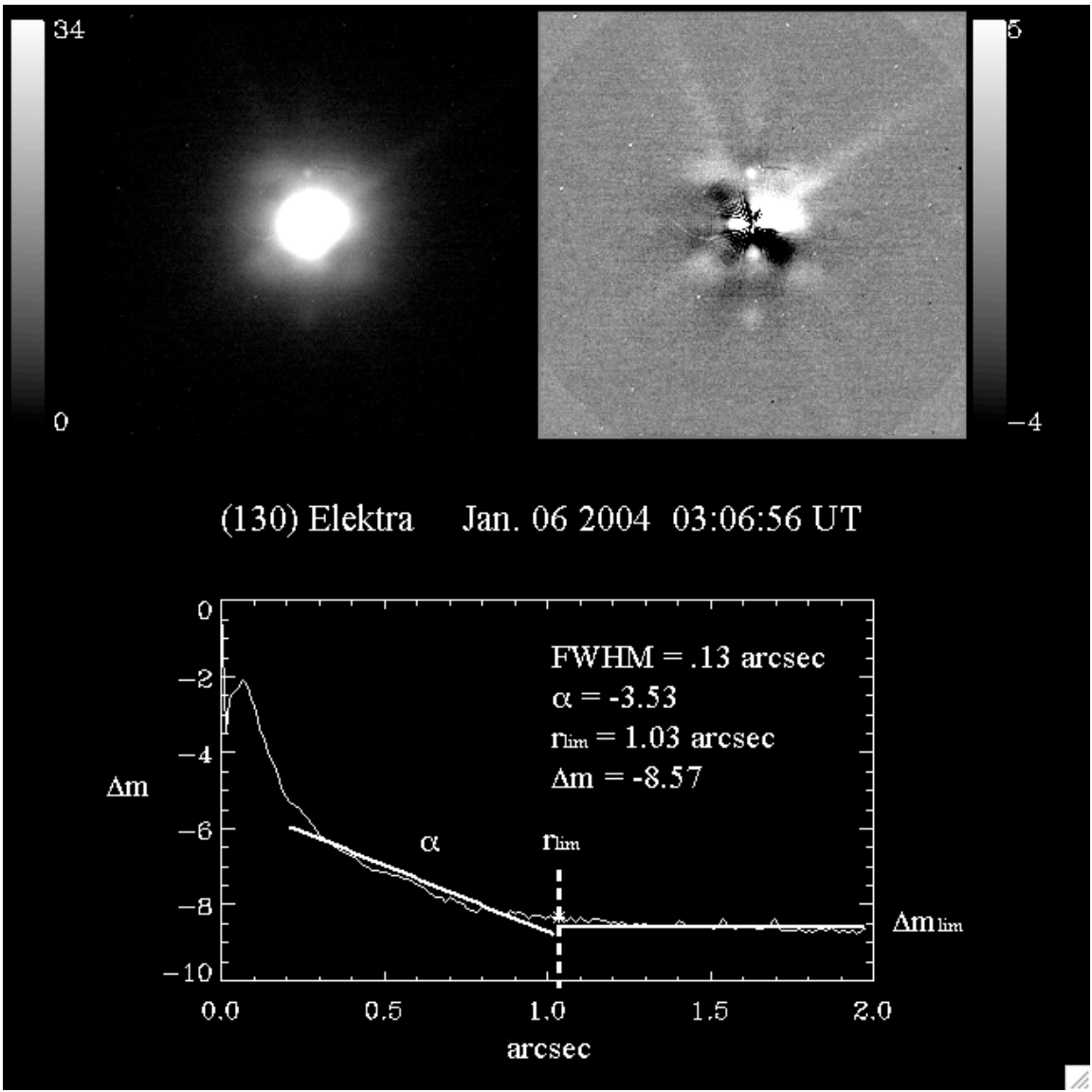

(130) Elektra    Jan. 06 2004   03:06:56 UT

FWHM = .13 arcsec
$\alpha = -3.53$
$r_{lim} = 1.03$ arcsec
$\Delta m = -8.57$

**Figure 1b:** Search for moonlets around (283) Emma for Dec. 28 2004 observations. Characteristics of the detection profile for all Emma observations can be found in Table 3a.

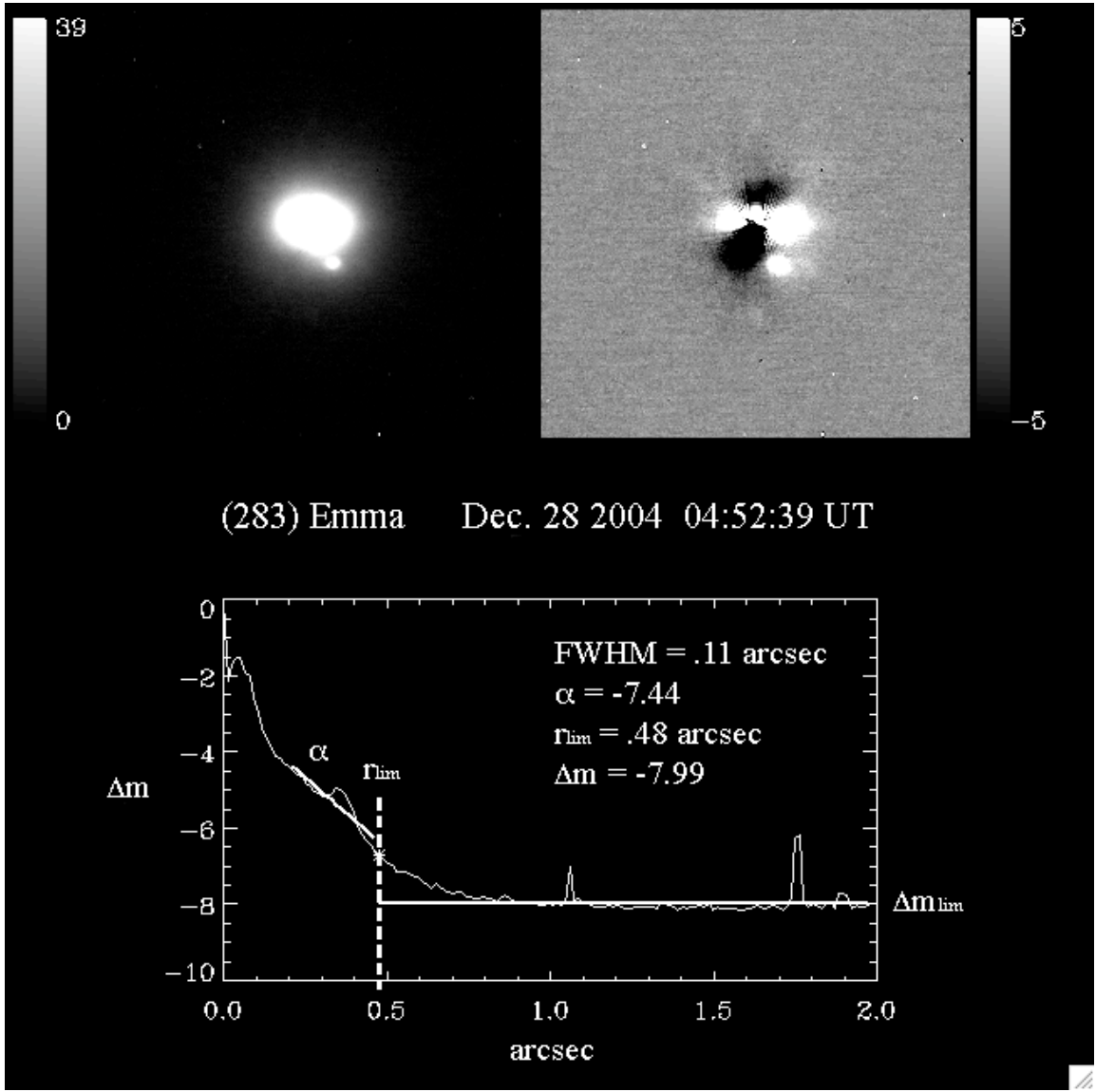

**Figure 1c:** Search for moonlets around (379) Huenna for Feb. 4 2005. Characteristics of the detection profile for all Huenna observations can be found in Table 3b.

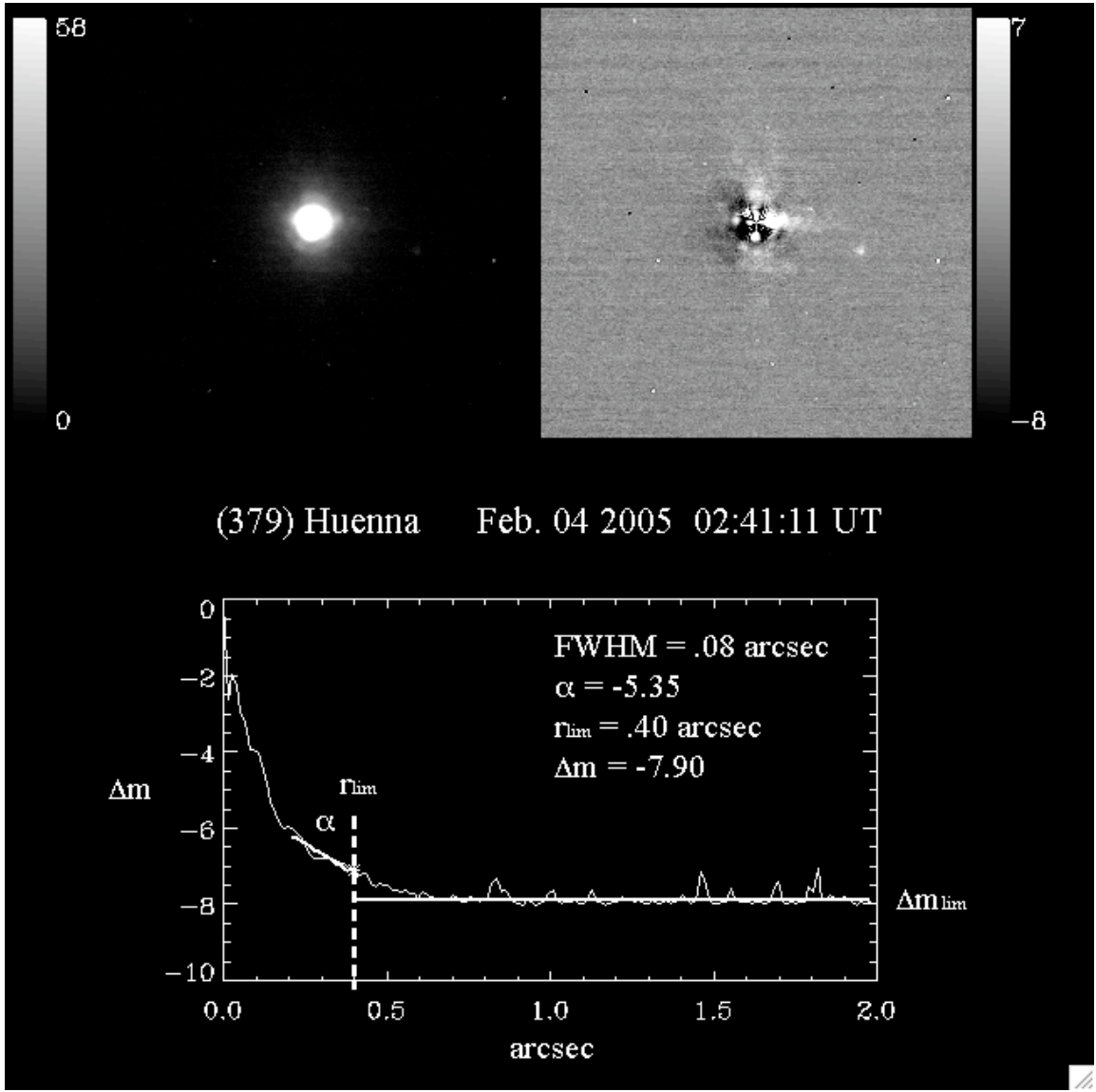

**Figure 1d:** Search for moonlets around (3749) Balam for Dec. 07 2004. Characteristics of the detection profile for all Balam observations can be found in Table 3b.

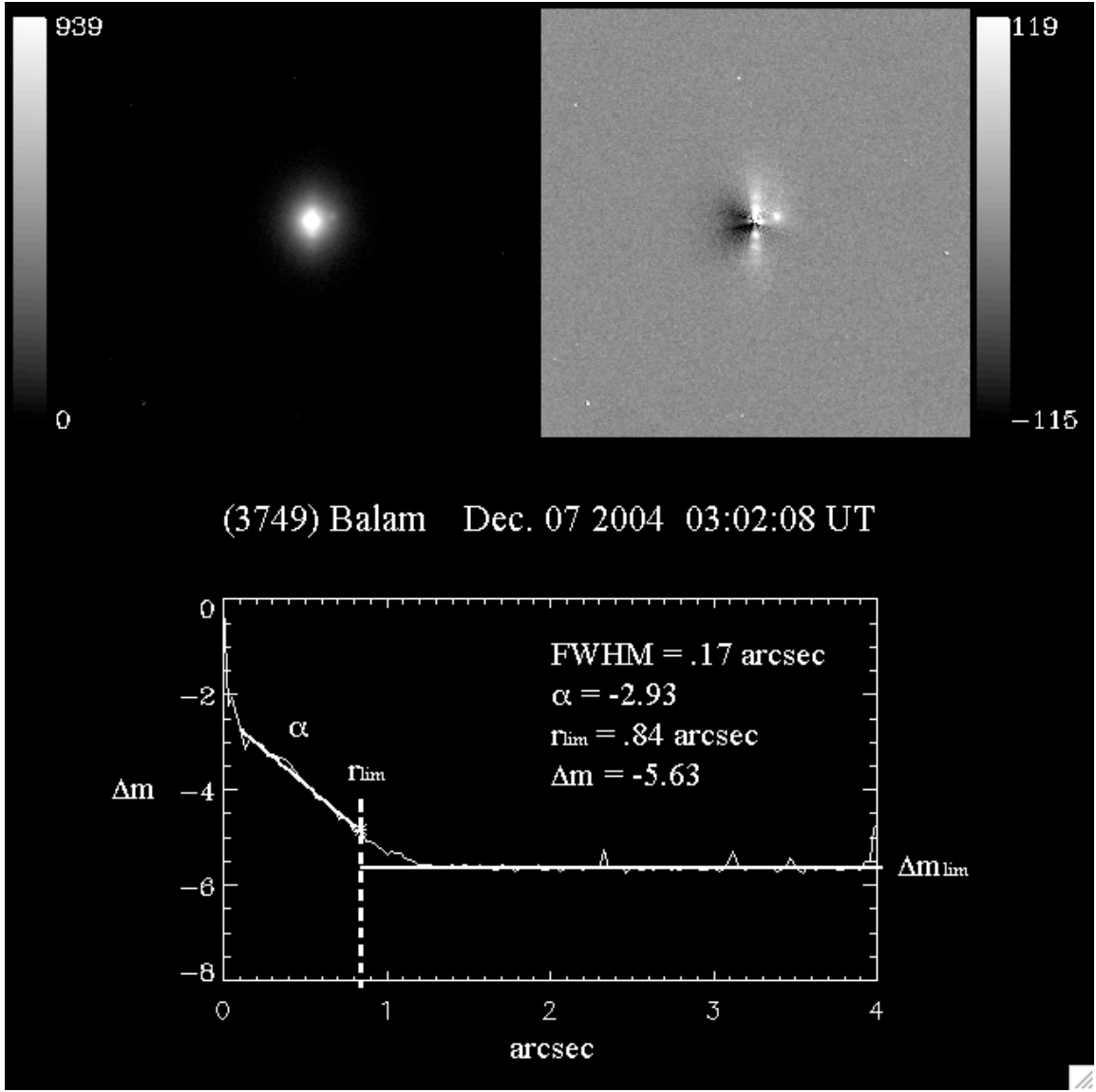

**Figure 2:** Shape and orientation of (130) Elektra and comparison with Durech *et al.* (2007) 3D-shape model (pole I), and its almost symmetrical solution determined from the moonlet orbit analysis (pole II, see Table 6). The apparent shape of (130) Elektra's primary (middle panel) is in agreement with the pole I model, implying that the almost symmetrical solution should be discarded. The apparent diameter of the primary varies because of different pixel scale between Gemini, VLT and Keck telescope NIR camera and the distance between the asteroid and Earth. A quantitative analysis between the pole I appearance model and the observations is included in Table 4a.

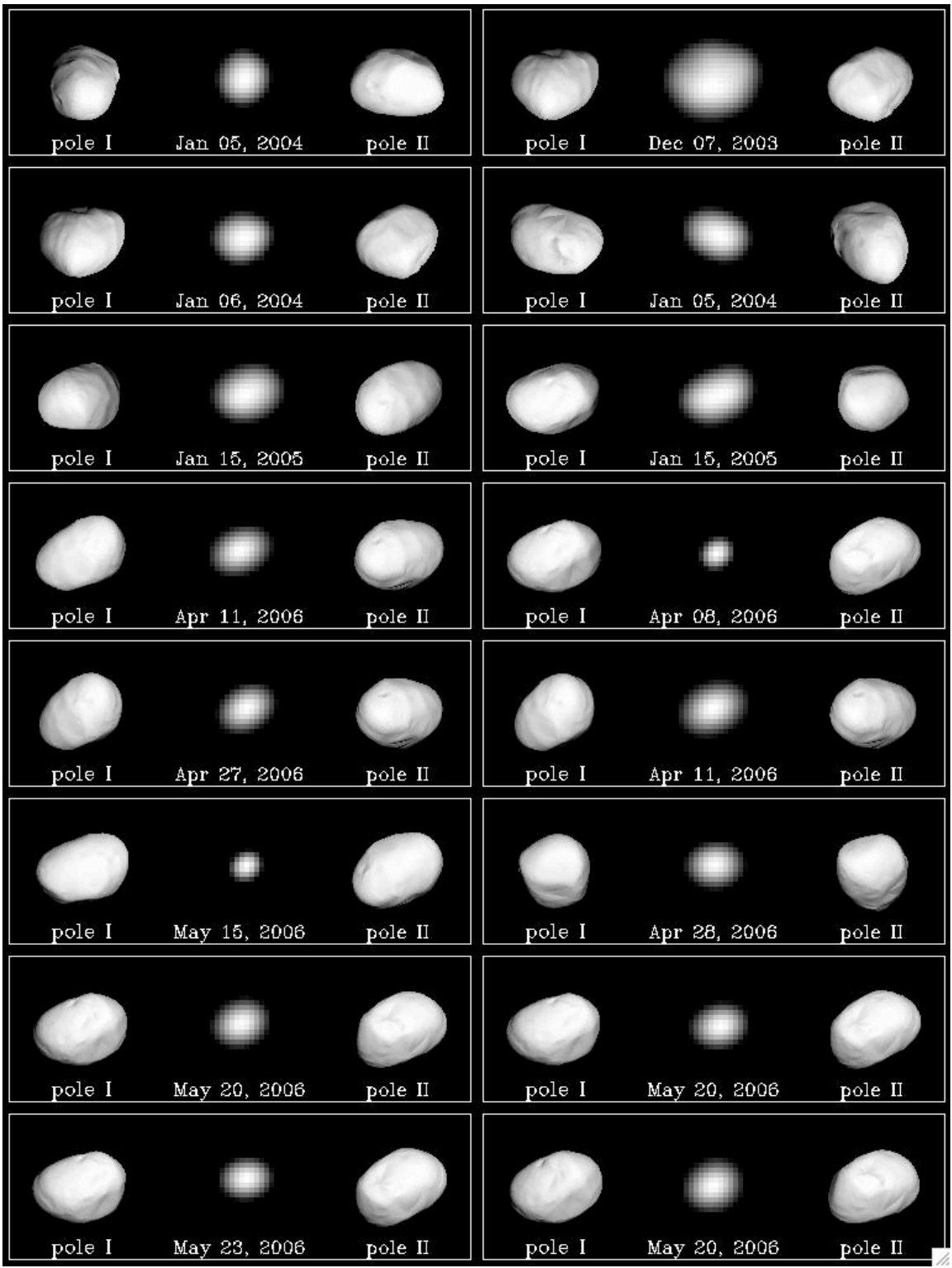

**Figure 3a**: [left] The apparent orbit of (130) Elektra's companion projected on the plane-of-sky. [right] Measured astrometric positions (crosses) from Table 5a and positions from our model (dots) are displayed. The solid lines represent the portion of the orbit in the foreground; the dashed line is in the background. The radial dashed line indicates the position of the pericenter.

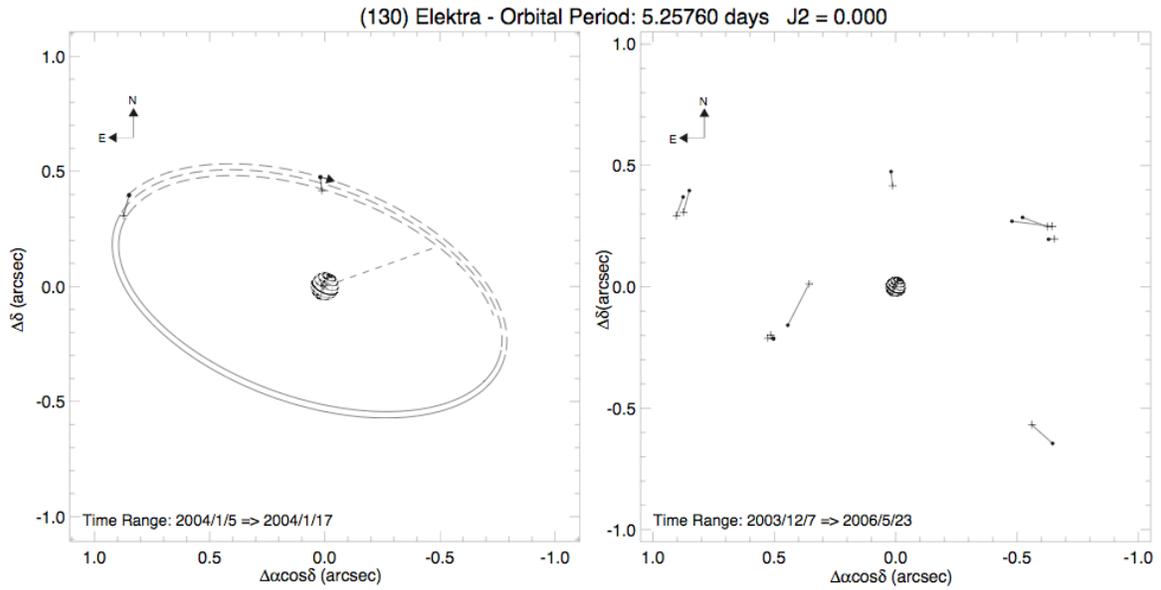

**Figure 3b:** The apparent orbit of (283) Emma's companion projected on the plane-of-sky.

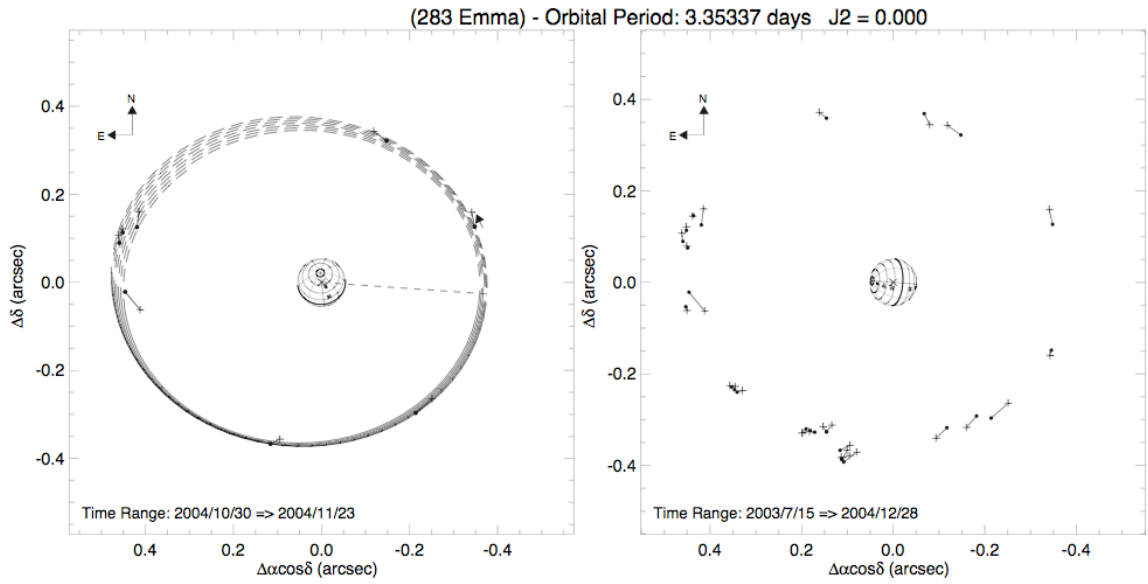

**Figure 3c:** The apparent orbit of (379) Huenna's companion projected on the plane-of-sky.

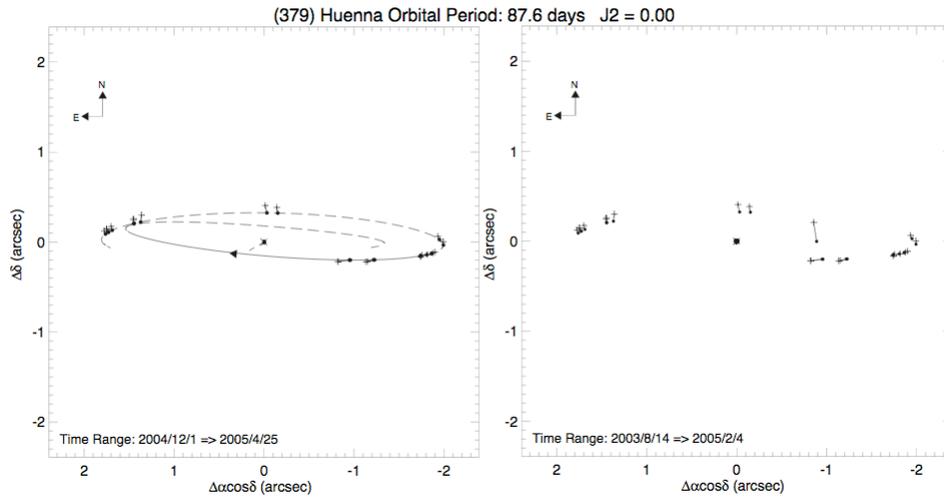

**Figure 3d:** The apparent orbit of (3749) Balam's companion projected on the plane-of-sky.

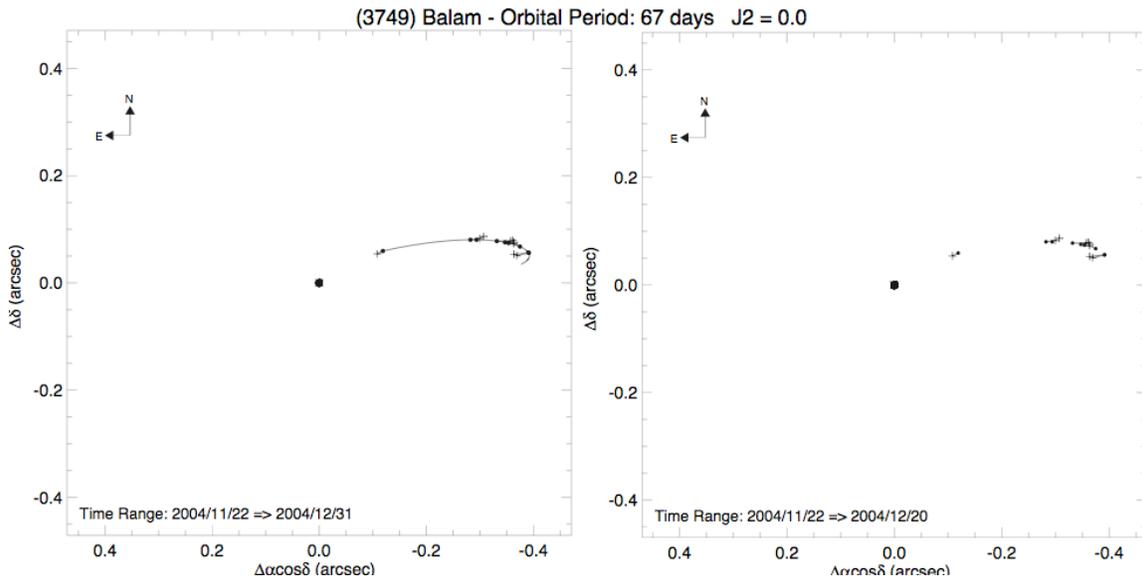

**Figure 4:** Evolution of binary asteroid mutual orbits due to tidal dissipation. A binary system with characteristics placing it above the synchronous stability limit (in bolt) will not evolve due to tidal effect. Similarly-sized binary systems, such as (90) Antiope, are located in this region (Descamps *et al.* 2007). Below the excitation limit curve the satellite of an asteroid will have its orbit excited by the tides. This limit was drawn under the assumption that the moonlet and the primary have the same coefficient of dissipation and bulk density which is highly unrealistic. However, Emma and Elektra binary systems are both located in this region and for both of them, their satellite has a significant eccentricity (~0.1). The almost-vertical dash lines define the timescale for the tides to act on the binary system. They were drawn assuming a density of 1.1 g/cm$^3$ for the primary and secondary and a product of rigidity and specific dissipation parameter μQ ~10$^{10}$, a possible value for a rubble-pile asteroid. In this case, Emma binary system appears quite young (~10 Myr) whereas Elektra is fairly old (~4.5 Byr).

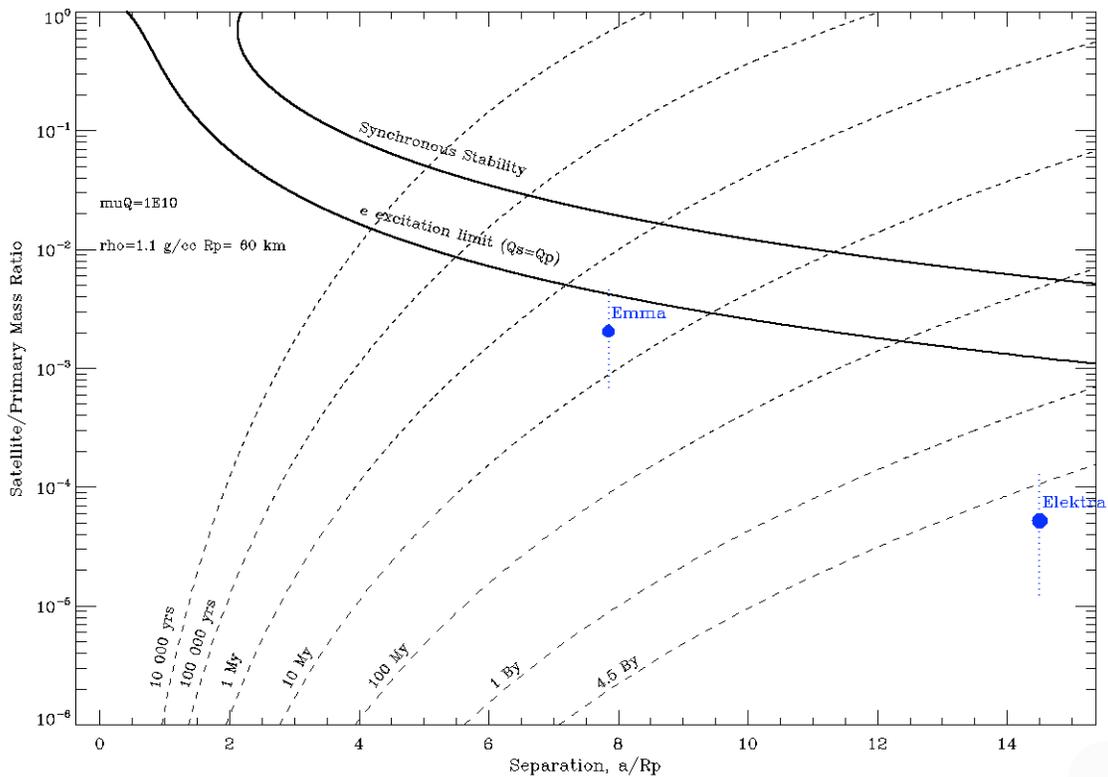